\documentclass[fleqn,usenatbib]{mnras}

\usepackage{newtxtext,newtxmath}
\usepackage[T1]{fontenc}

\DeclareRobustCommand{\VAN}[3]{#2}
\let\VANthebibliography\thebibliography
\def\thebibliography{\DeclareRobustCommand{\VAN}[3]{##3}\VANthebibliography}

\usepackage{graphicx}	% Including figure files
\usepackage{amsmath}	% Advanced maths commands
\usepackage{tikz}       % For Orchid-ID thing
\usepackage{xcolor}
\usepackage{hyperref}

\AtBeginDocument{%
  \DeclareRobustCommand{\J18}{\hyperlink{cite.jerkstrand2018emission}{J18}}%
}
\makeatother
\newcommand\kms[1]{{$#1\,\text{km}\,\text{s}^{-1}$}}

\newcommand\txtred[1]{{\color{black}#1}}  % To red-text things <-- CHANGES following referee report!
\newcommand{\angstrom}{\mbox{\normalfont\AA}}
\definecolor{lime}{HTML}{A6CE39}
\DeclareRobustCommand{\orcidicon}{%
    \begin{tikzpicture}
    \draw[lime, fill=lime] (0,0) 
    circle [radius=0.16] 
    node[white] {{\fontfamily{qag}\selectfont \tiny ID}};
    \draw[white, fill=white] (-0.0625,0.095) 
    circle [radius=0.007];
    \end{tikzpicture}
    \hspace{-2mm}
}
\newcommand{\orchidJanka}{\href{https://orcid.org/0000-0002-0831-3330}{\orcidicon}}
\newcommand{\orchidJerkstrand}{\href{https://orcid.org/0000-0001-8005-4030}{\orcidicon}}
\newcommand{\orchidKresse}{\href{https://orcid.org/0000-0003-1120-2559}{\orcidicon}}
\newcommand{\orchidVanBaal}{\href{https://orcid.org/0009-0001-3767-942X}{\orcidicon}}

\title[3D NLTE radiative transfer]{Emission line models for the lowest mass core-collapse supernovae -- II. 3D NLTE radiative transfer modelling of a $9.0\,M_\odot$ neutrino-driven explosion}  % ?Temporary working title?
\author[B. van Baal et al.]{
Bart F. A. van Baal\orchidVanBaal,$^{1}$\thanks{E-mail: barteld.vbaal@astro.su.se}
Anders Jerkstrand\orchidJerkstrand,$^{1}$
Daniel Kresse\orchidKresse,$^{2}$%
and Hans-Thomas Janka\orchidJanka$^{2}$
\\
$^{1}$The Oskar Klein Centre, Department of Astronomy, Stockholm University, AlbaNova, Se-10691 Stockholm, Sweden\\
$^{2}$Max Planck Institute for Astrophysics, Karl-Schwarzschild-Str 1, D-85748 Garching, Germany\\
}

\date{Accepted XXX. Received YYY; in original form ZZZ}

\pubyear{\the\year{}}

\begin{document}
\label{firstpage}
\pagerange{\pageref{firstpage}--\pageref{lastpage}}
\maketitle

\begin{abstract}
%% 28-04-26, 243 words (i.e. at the limit)
The nebular phase of a supernova (SN) occurs several months to years after the explosion, \txtred{with asymmetries created by the explosion encoded into the line profiles of the emission lines. To} make accurate predictions for these line profiles, Non-Local Thermodynamic Equilibrium (NLTE) radiative transfer calculations need to be carried out. In this work, we use \texttt{ExTraSS} (EXplosive TRAnsient Spectral Simulator) -- which was recently upgraded into a full 3D NLTE radiative transfer code (including photoionization and line-by-line transfer effects) -- to \txtred{perform} such calculations. \texttt{ExTraSS} is applied to a 3D explosion model of a $9.0\,M_\odot$ H-rich progenitor\txtred{,} evolved into the homologous phase. Synthetic spectra are computed and lines from different elements are studied for varying viewing angles. \txtred{Line profile properties strongly correlate with a primary Ni plume in the ejecta.} The model spectra are compared against observations of SN~1997D and SN~2016bkv. The model \txtred{can create good} line profile matches for both SNe, and reasonable luminosity matches for He, C, O, and Mg lines for SN~1997D -- however H$\alpha$ and Fe~I lines are too strong. \txtred{Key diagnostic lines of low-mass core-collapse SNe (CCSNe), e.g. differentiating Fe CCSNe from electron capture SNe, are upheld from 1D to 3D. However, both line profiles and line luminosities differ in 3D across viewing angles, enabling the possibility of detailed comparisons to observed spectra to infer asymmetries imprinted by the explosion. We show that even the fastest $^{56}$Ni is traceable in nebular phase line profiles.}
% It should be a single paragraph not more than 250 words (200 words for Letters).
% No references should appear in the abstract.
\end{abstract}


%%%%%%%%%%%%%%%%%%%%%%%%%%%%%%%%%%%%%%%%%%%%%%%%%%

%%%%%%%%%%%%%%%%% BODY OF PAPER %%%%%%%%%%%%%%%%%%

\section{Introduction}

Massive stars ($M_\mathrm{ZAMS}\gtrsim8\,M_\odot$) end their lives in a core-collapse supernova (CCSN) event \citep{heger2003massive,jerkstrand2026core}. During these explosions, the core of the star forms either a neutron star or black hole, while the rest of the star gets ejected and enriches the interstellar medium with the elements synthesized during both hydrostatic and explosive burning \citep{woosley1995evolution,arnett1996supernovae,woosley2002evolution,limongi2003evolution,ceverino2009role}. Given a Salpeter initial mass function, stars in the mass range of $M_\mathrm{ZAMS} = 8-12\,M_\odot$ should account for $\sim40\,\%$ of all CCSNe, or perhaps even as high as $\sim60\,\%$ if a significant fraction of stars with $M_\mathrm{ZAMS} > 20\,M_\odot$ collapse directly to black holes without producing a SN event \citep{sukhbold2016corecollapse}. However, the core structure of the stars in this mass range is starkly different from the structure of stars with $M_\mathrm{ZAMS}>12\,M_\odot$ \citep{sukhbold2016corecollapse}, leading to a lower compactness \citep{oconnor2011black} and making it easier to explode such stars -- even in 1D neutrino\txtred{-driven CCSN} simulations, explosions can be achieved \citep{kitaura2006explosions,fischer2010protoneutron,melson2015neutrino,radice2017electron}.

However, it is unclear if all the stars in the $M_\mathrm{ZAMS} = 8-12\,M_\odot$ range end their lives as CCSNe after forming an iron core, or if some explode as electron-capture SNe (ECSNe) upon the formation of the oxygen-neon-magnesium core, or end up as white dwarf \citep{nomoto1984evolution,nomoto1987evolution,jones2013advanced,woosley2015remarkable,doherty2015super,doherty2017super,jerkstrand2018emission,kozyreva2021synthetic}. The evolution of the lowest-mass progenitors is difficult to model, owing to thermal pulses and flashes and degeneracy effects \citep[see e.g.][]{miyaji1980supernova,jones2016electroncapture}, creating a sparsity of models. The models which do exist indicate that ECSNe, as well as the most marginal cases of iron core collapse, have low explosion energies and expansion velocities, low $^{56}$Ni ejecta masses, and low luminosities \citep{janka2012explosion,eldridge2019supernova,burrows2019three,stockinger2020three,burrows2021corecollapse,sandoval2021three}. 

Observationally, a few stellar explosions have been suggested to represent ECSNe -- including the historical SN~1054 \citep[the Crab, see e.g.][but see also \citealt{gessner2018hydrodynamical,temim2024dissecting}]{nomoto1982crab,hillebrandt1982exploding,tominaga2013supernova,smith2013crab}. SN~2016bkv was tentatively suggested by \citet{hosseinzadeh2018short} to be of ECSN origin due to its peculiar nebular phase spectra and a comparison against models from \citet[][\J18 hereafter]{jerkstrand2018emission}, although the inferred $^{56}$Ni mass ($0.02\,M_\odot$) seems to conflict with the low yields expected for ECSNe ($\lesssim 5\times 10^{-3}\,M_\odot$). SN~2018zd was also suggested to be an ECSN \citep[][but see also \citealt{callis2021low}]{hiramatsu2021electron}, although it was quite luminous with a peak brightness of $-18.40\,\pm\,0.60\,$mag.

%Typically, the properties of (lookalike-)ECSNe would classify them as a subset of Type IIP SNe, 
A promising observational class to match the $8-12$ $M_\odot$ range are the so-called low-luminosity (sometimes "subluminous") Type IIP SNe \citep[LLIIP][]{pastorello2004low,spiro2014low,mullerbravo2020low,dastidar2025sn2018is,das2025low}. LLIIP are rare ($\sim5-10\,\%$ of all Type II SNe, \citealt{pastorello2004low,das2025low}) which, if these SNe originate from $M_\mathrm{ZAMS} = 8-12\,M_\odot$ stars, indicates we are missing a large fraction of these \citep[see e.g.][]{horiuchi2011cosmic,jencson2019SPIRITS}, or that the mass range is significantly smaller. This could be due to their intrinsic faintness or because a significant fraction of them could be obscured by dust \citep{jencson2019SPIRITS}. Pre-explosion images of the progenitors of SN~2005cs \citep{maund2005progenitor,li2006identification}, SN~2008bk \citep{mattila2008VLT} and SN~2022acko \citep{vandyk2023identifying} indicate that these LLIIP have low-mass progenitors ($M_\text{prog}\lesssim11\,M_\odot$). Furthermore they have low $^{56}$Ni yields ($\sim0.005\,M_\odot$), about an order of magnitude lower than usual Type IIP SNe \citep{spiro2014low}, as well as low expansion velocities ($\sim\,1300-2500\,\text{km}\,\text{s}^{-1}$ at 50 days after explosion, \citealt{das2025low}, using M$_{r,\text{peak}}>‑16\,$magnitude as cut off magnitude).

Only a few LLIIP have been detected in the nebular phase. The first of these is SN~1997D, which was originally discovered by \citet{demello1997supernova}, with nebular phase spectra by \citet{turatto1998peculiar} and \citet{benetti2001fading}. At that time, SN~1997D was the least luminous and energetic Type II SN discovered, with expansion velocities of order \kms{1000}. It wasn't clear if this SN originated from a low-mass progenitor with a low explosion energy, or from a more massive star with significant fallback material. Initially, this latter case was favoured as a few more LLIIP were detected \citep{pastorello2004low}. The discovery and detailed study of SN~2005cs \citep{pastorello2006sn2005cs,pastorello2009sn2005cs} turned this around towards the low-mass progenitor scenario, as the progenitor star could be identified in archival Hubble Space Telescope images \citep{maund2005progenitor} and was estimated to be $7-13\,M_\odot$. As more and more LLIIP were discovered \citep[e.g. SN~2009md by][]{fraser2011SN2009md}, the low-mass progenitor scenario became the favoured explanation, especially when transients with properties in between the regular Type IIP and LLIIP started appearing and also had low(er) mass estimates \citep[e.g. SN~2009N by][]{takats2014sn2009n}. In the last decade, several other LLIIP have been detected in the nebular phase, including SN~2016aqf \citep{mullerbravo2020low}, SN~2016bkv \citep{hosseinzadeh2018short,nakaoka2018low}, SN~2018lab \citep{pearson2023circumstellar}, and SN~2020cxd \citep{yang2021low,kozyreva2022low,valerin2022low}. %% NOTE TO SELF - a LOT of these seem to have early CSM interactions... so mass loss very late prior to explosion (or pulsation?), so perhaps that can 'correct' the explosion energy versus ejecta velocity?

Light curve simulations also favour the low-mass progenitors for LLIIP \citep{fraser2011SN2009md,pumo2017radiation,kozyreva2021synthetic}, as does the evolutionary numerical simulation of a low mass red supergiant for SN~2008bk \citep{lisakov2017study}. Additionally, 1D nebular phase spectroscopy \txtred{(\J18, \citealt{dessart2021explosion})} indicates good matches to several of the LLIIP for $M_\mathrm{ZAMS} = 8-12\,M_\odot$ models. 

Although evolutionary models of stars in the $M_\mathrm{ZAMS}=8-12\,M_\odot$ range are rare and CCSN explosions can be achieved in 1D for part of this mass range, a few 3D simulations have been carried out to shock breakout and beyond \citep{stockinger2020three,sandoval2021three,vartanyan2025simulated}. \citet{stockinger2020three} exploded three different low-mass progenitors (in the range $8.8-9.6\,M_\odot$) and evolved them through shock breakout until fallback was completed. Their s9.0 model (a solar-metallicity star with $M_\mathrm{ZAMS} = 9\,M_\odot$, from \citealt{sukhbold2016corecollapse}) is of particular interest, as it achieves a degree of mixing similar to that of more massive stars \citep[e.g. the models from][]{wongwathanarat2015three} despite the much lower explosion energy, and has significantly asymmetric ejecta with the fastest $^{56}$Ni being ejected with velocities up to $1400\,\text{km}\,\text{s}^{-1}$. The s9.0 model from \citet{stockinger2020three} was modelled for 19.74 days post-bounce, at which point the ejecta is roughly homologously expanding (with shock breakout beginning at $\sim$2.2 days and completing at $\sim$3.3 days). 

% \daniel{Should we also give credit to recent progress in 3D spectral synthesis modeling for SNe Ia, e.g., Sim2007, Kromer+Sim2009, Pollin+2024,2025? (Pollin+2025 also computed 3D NLTE nebular-phase spectra. I cannot comment on their quality; just came across this recent work and wanted to point it out in case you didn't already know it.)}
The study of spectra of 3D CCSN models \txtred{has begun} recently. \citet{jerkstrand2020properties} developed a new platform for 3D spectral synthesis, and applied it to study $\gamma$-ray lines and approximate optical and NIR lines, for a suite of Type IIP explosion simulations of $M_\mathrm{ZAMS}=15-20\,M_\odot$ progenitors. This initial platform could do ray tracing as well as Compton scattering on a 3D spherical coordinate system. However, it lacked capacity to compute temperature, ionization, and excitation. This major step was taken by \citet{vanbaal2023modelling}, implementing the microphysics and atomic data of the 1D code \texttt{SUMO}, with several improvements, into a much improved 3D code, which was also named \texttt{ExTraSS} (EXplosive TRAnsient Spectral Simulator). This code version was used to study stripped-envelope SNe\footnote{Stars which have lost their H-rich (and sometimes also He-rich) envelope, leading to SNe without spectral signatures of H (and He).} \citep{vanbaal2023modelling,vanbaal2024diagnostics}. For such SNe, radiative transfer effects could largely be approximated to only occur locally, due to their high expansion velocities and lower ejecta masses, thus this code version used an on-the-spot treatment for photoionization.

Recently, \texttt{ExTraSS} has undergone a second major upgrade \citep{vanbaal2025ExTraSS}, adding in radiative transfer for photoexcitation and photoionization. With this capacity, it becomes possible to study also Type II SNe, where these transfer effects are important also in the nebular phase. In particular, it enables to process these new 3D hydrodynamic simulations of low-mass progenitors. This in turn opens the possibility to move from 1D models \txtred{(\J18, \citealt{dessart2021explosion})} to obtain more accurate predictions. For example, 1D models predicted certain distinguishing features between Fe CCSNe and ECSNe, but the artificial shell structure in 1D gives uncertainty to whether these are seen also in 3D. The study of these features is one of the main goals of this paper. We also aim to determine the viewing angle effects on line luminosities and line profiles, and compare the new models to recent data sets.

The paper is structured as follows: in Section~\ref{sec:methods} we briefly describe the radiative transfer upgrade to \texttt{ExTraSS} (full details in \citealt{vanbaal2025ExTraSS}), as well as the s9.0 model from \citet{stockinger2020three}. In Section~\ref{sec:results} we present our results, which we discuss in Section~\ref{sec:discussion} before we summarize our findings in Section~\ref{sec:conclusion}.

% NOTE TO SELF; I should add a cover letter when submitting to MNRAS to outline the link between the two papers. I do not think I can link them on arXiv submission (but, I can use the arXiv links when resubmitting)

\section{Methods} \label{sec:methods}
\subsection{Radiative transfer and \texttt{ExTraSS}} \label{ssec:3DRadTrans}
In this work, we apply the \texttt{ExTraSS} code (see also \citealt{vanbaal2025ExTraSS}, for a full description) which has been upgraded with a new radiative transfer treatment, which we summarize here.

We compute the radiative transfer by considering photoionization and lines. We replace the ``on-the-spot'' photoionization treatment of \citet{vanbaal2024diagnostics} with photoionization rates computed from the radiation field, and now also compute photoexcitation rates. As the radiation field is both generated by the level populations in each cell, and interacts with them through photoionization and photoexcitation, we also need to iterate between the level population solver \citep[introduced in][]{vanbaal2023modelling} and the new radiative transport module. Such $\Lambda$-iteration schemes have good convergence properties for the Sobolev approximation at intermediate optical depths \citep[see][for a recent review]{jerkstrand2025spectral}. 

Within each cell, the generated emission is binned by wavelength ($400-25000\,\angstrom$, with logarithmic step sizes of $0.1\%$) to create photon packets. Emission blueward of $1$\txtred{$5$}$000\,\angstrom$ is treated with the new radiative transport, while the emission between $1$\txtred{$5$}$000-25000\,\angstrom$ is treated as before, under an optically thin approximation\footnote{This choice reduces the number of bins for which radiative transport is used to \txtred{3660}, while it would be 4170 if the transport treatment was extended to $25000\,\angstrom$. As longer wavelengths are less impacted by radiative transport effects, this gives a good compromise between computational cost and accuracy. Escaped emission further redwards of $25000\,\angstrom$ is accounted for in the convergence check.}. The emission mechanisms considered are bound-bound emission, recombination emission, and two-photon emission for neutral H \citep{nussbaumer1984hydrogenic} and He \citep{li1995hetwophoton}.

We use a ray-tracing technique to follow the path of the photon packets from their point of emission until they either escape the grid or less than $10^{-6}$ of their starting energy remains. To limit memory usage of tracking the photoexcitation rates we adapt a domain decomposition scheme (see below) similar to \citet{brunner2009efficient}, such that each node only stores rates for its own section while still having an efficient global transport. Rays are emitted towards all viewing angles; we use $20\times20$ viewers \citep[as in][]{vanbaal2023modelling,vanbaal2024diagnostics} equally spread in the polar and azimuthal angles.

The ray tracing is done in a similar manner as described in \citet{jerkstrand2020properties}, but with modifications to the optical depths and the division of emission into the energy packets ($j_i$ in \citealt{jerkstrand2020properties}). The first change is to calculate the optical depths with respect to photoionization, instead of Compton scattering, while the energy packet subdivision is altered to adjust for the solid angle of the viewing direction ($\Delta\Omega_k$ in \citealt{jerkstrand2020properties}). This is done because the viewing directions are not spread isotropically\footnote{With a setup of $20\times20$ viewing angles for the polar$\,\times\,$equatorial angles, viewing angles closer to the north/south poles are packed more tightly and cover a smaller patch of the sky.}, which means without such a correction the emission would not be isotropic as it should be. Therefore, the final output spectra are now also corrected for this solid angle.

In addition to the radiative transfer code expansion, \texttt{ExTraSS} has also been updated to calculate energy deposition for non-thermal excitations \citep[using the method of][]{kozma1992gamma}, to use specific recombination rates for \txtred{H~I}, O~I, \txtred{Mg~I,} Fe~I and Fe~II, and with better photoionization cross section calculations (using data from \citealt{verner1996atomic}) \txtred{alongside an adjustment for the number of levels used to compute photoionization, variable per ion}. Internal collisional rates for $nl$-levels in hydrogen are also calculated \citep{pengelly1964recombination,brocklehurst1971calculations} in more detail.

In order to manage the memory load of 3D NLTE radiative transfer, \texttt{ExTraSS} has been adapted to a domain decomposition scheme which is inspired by \citet{brunner2009efficient}. The goal of this technique is to reduce memory load on the system at the cost of some communication overhead. This communication overhead can be reduced by properly splitting the full domain into smaller ones, which in our case means $\phi$-based decomposition, i.e. each node takes an equal part of the number of azimuthal slices to make a domain. Through (intra-node) memory sharing, all critical information on the domain can be made available for each ray on that domain, further limiting memory usage. Verification of the robustness of this scheme is given in \citet{vanbaal2025ExTraSS}, alongside a more detailed explanation of the implementation.

\subsection{Explosion model} \label{ssec:model} % used for referring to this section from elsewhere
\begin{figure*}  %% Not sure if I like this as column-width instead of full-page
    \centering
    \includegraphics[width=\linewidth]{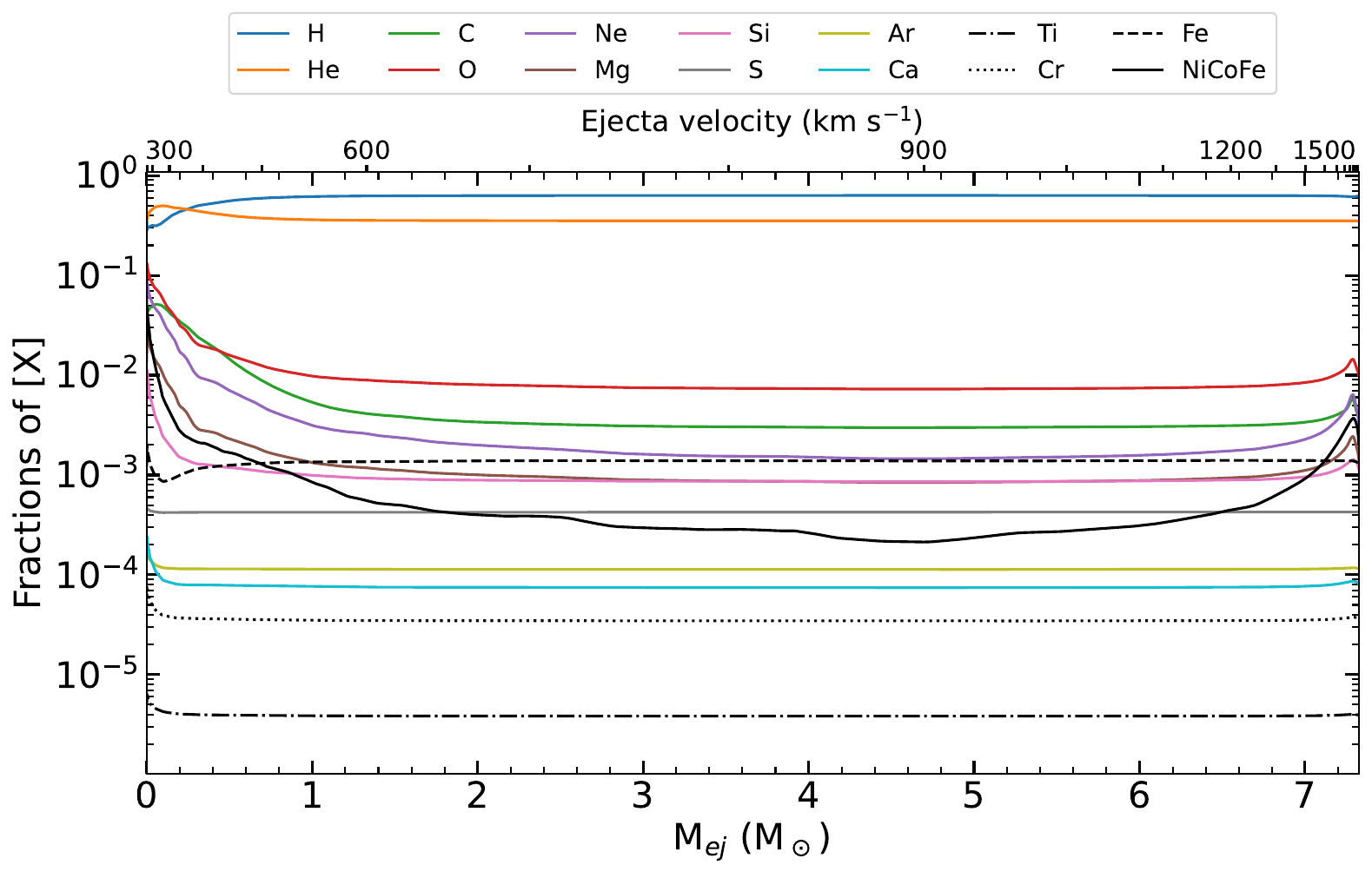}
    \caption{Mass fractions of the angle-averaged ejecta for s9.0. \txtred{Each of the $\alpha-$nuclei present in the \texttt{Prometheus} nuclear network is shown, with the `NiCoFe' label indicating the explosively produced radioactive $^{56}$Ni and its daughter isotopes $^{56}$Co and $^{56}$Fe, while the dashed black `Fe' indicates the stable, pre-existing iron from the progenitor star.} The neutron star mass ($1.355\,M_\odot$, \citealt{janka2024interplay}) is excluded in this plot. The top axis indicates the velocity of the ejecta. The total ejecta masses for each element are also listed in Table~\ref{tab:StockMass}. \txtred{3D renderings}, up to 2.8\,d, of the $^{56}$Ni \txtred{are} shown in \citet[Figure~23\txtred{; see also \citealt{kozyreva2022low}, Figure~1}]{stockinger2020three}.}
    \label{fig:composition_stock}
\end{figure*}
The original stellar evolution of the s9.0 model was done by \citet{sukhbold2016corecollapse}, where it was the least massive model they considered as it is the lowest mass star which undergoes Fe-CCSN in the \texttt{KEPLER} code \citep{weaver1978presupernoa,woosley2015remarkable}. The initial 3D explosion simulation of s9.0 was done by \citet{melson2020resolution} in \texttt{Prometheus-VERTEX} \citep{rampp2002radiation,buras2006two} with detailed neutrino physics. After $\sim500\,$ms (since bounce), the neutrino transport was replaced by a simplified neutrino heating and cooling scheme, for computational reasons. At $3.140\,$seconds after bounce, when the explosion energy has reached its asymptotic limit ($E_\text{expl}=0.54\times10^{50}\,$erg, see Table 2 in \citealt{janka2024interplay}), the model was mapped onto a new grid with \texttt{Prometheus-HotB} \citep[similar to][]{wongwathanarat2015three}. From this mapping point \citet{stockinger2020three} ran the simulation until $\sim19.74\,$days post-bounce. Decay of radioactive nickel is included (\citealt{stockinger2020three}, section 3.5, see also \citealt{gabler2021infancy}) as internal energy source. A more thorough description of the \txtred{\texttt{P-VERTEX} and \texttt{P-HotB}} codes is given in \citet{stockinger2020three}. The final kick velocity of the neutron star is \kms{57} \citep{janka2024interplay} and is dominated by the \txtred{neutrino-induced kick, which is caused by anisotropic neutrino emission, rather than the hydrodynamical one connected to asymmetric mass ejection}.

\looseness-1The data of the s9.0 model was mapped from the Yin-Yang grid used for the simulation onto a spherical polar grid with $2268\times90\times180$ for $N_r\times N_\theta \times N_\phi$, and rotated around the x-axis by 60 degrees (such that the major $^{56}$Ni plume points in a direction roughly perpendicular to the polar axis of the spherical grid). The innermost radii (up to $v_\text{ej}\,\lesssim\,25\,\text{km}\,\text{s}^{-1}$) contain very little mass, as do cells with $v_\text{ej}\,\gtrsim\,2200\,\text{km}\,\text{s}^{-1}$. These radii are therefore not used in our input model, to keep the grid as small as possible. The angle-averaged compositional structure of the ejecta is shown in Figure~\ref{fig:composition_stock}, and the elemental contribution to the ejecta is summarized in Table~\ref{tab:StockMass}.
The remaining grid is downsized to $N_r\times N_\theta \times N_\phi=38\times15\times30$, which gives enough resolution to capture the 3D asymmetries while limiting the computational costs. This results in a final $\Delta r/r$ of $11\,\%$ for the radial steps, and an angular resolution of $12^\circ$ for the $\theta$ and $\phi$ grid. The grid is homologously extrapolated from the $19.74\,$day endpoint of \citet{stockinger2020three} to each modelled nebular phase epoch.

\begin{table}
 \centering
 \caption{Ejecta masses per element for the model, \txtred{at $400\,$days post-bounce}. All of the $^{56}$Ni (\txtred{$M_\text{Ni}=6.058\times10^{-3}\,M_\odot$}) has already decayed into $^{56}$Co and $^{56}$Fe. The total ejecta mass is \txtred{$7.32\,M_\odot$}.}
 \label{tab:StockMass}
 \setlength\tabcolsep{10pt}
 \begin{tabular*}{.5\linewidth}{l c} 
  \hline
  Element & Ejected Mass \\
          & ($M_\odot$) \\
  \hline
  \txtred{H}   & \txtred{$4.506$} \\
  \txtred{He}  & \txtred{$2.641$} \\
  \txtred{C}   & \txtred{$4.071\times 10^{-2}$} \\
  \txtred{O}   & \txtred{$7.401\times 10^{-2}$} \\
  \txtred{Ne}  & \txtred{$2.459\times 10^{-2}$} \\
  \txtred{Mg}  & \txtred{$1.016\times 10^{-2}$} \\
  \txtred{Si}  & \txtred{$7.229\times 10^{-3}$} \\
  \txtred{S}   & \txtred{$3.104\times 10^{-3}$} \\
  \txtred{Ar}  & \txtred{$8.342\times 10^{-4}$} \\
  \txtred{Ca}  & \txtred{$5.582\times 10^{-4}$} \\
  \txtred{Ti}  & \txtred{$2.837\times 10^{-5}$} \\
  \txtred{Cr}  & \txtred{$2.569\times 10^{-4}$} \\
  \txtred{Fe}  & \txtred{$1.555\times 10^{-2}$} \\
  \txtred{Co}  & \txtred{$2.149\times 10^{-4}$} \\
  \hline
 \end{tabular*}
\end{table}

\begin{figure}
    \centering
    \includegraphics[width=\linewidth]{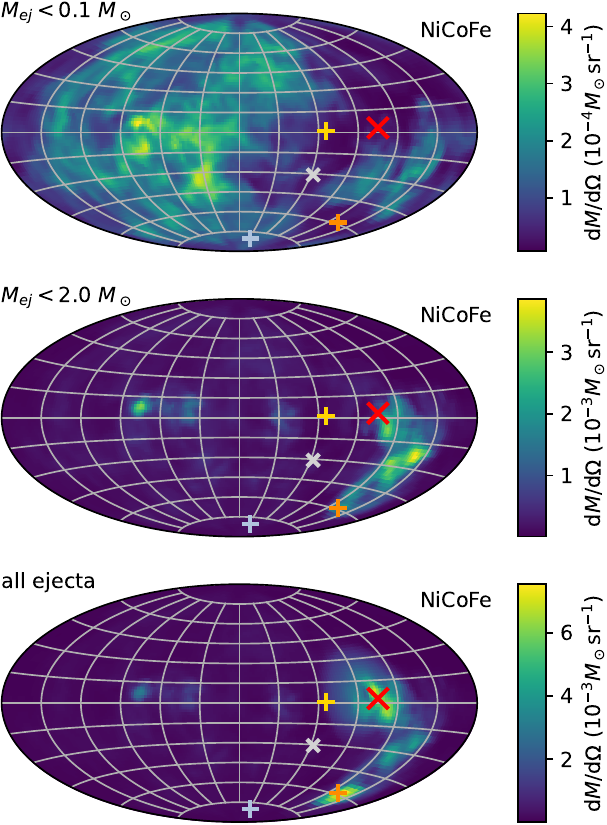}
    \caption{\txtred{Aitoff projections indicating the areas within the model where $^{56}$Ni and its daughter nuclei $^{56}$Co and $^{56}$Fe are located. The top panel shows the innermost $0.1\,M_\odot$ of the ejecta, the middle panel up to $2.0\,M_\odot$, and the final panel includes the whole ejecta mass (at 19.74 days post-explosion). In each panel, the red cross marks the direction towards the primary Ni plume, the orange plus towards a smaller secondary Ni plume, the light blue plus marks the hydrodynamical neutron star kick due to asymmetric mass ejection, the yellow plus the neutrino-induced kick due to anisotropic neutrino emission, and the gray cross the total (hydrodynamical+neutrino-induced) kick.}}
    \label{fig:Niplume-aitoff}
\end{figure}

%Aitoff projections of the NiCoFe mass distribution, dM/dOmega, at three different mass coordinates (0.1 Msun, 2.0 Msun, and including the entire ejecta). I computed the mass distributions as follows:
% dM/dOmega = \int_0^R density * X_NiCoFe * r^2 dr,
%where R = R(M_ej) is the radius that corresponds to the specified mass coordinate. The red crosses in the panels mark the position of the major nickel-rich plume
\txtred{To better outline the location of the major $^{56}$Ni plumes in the model, in Figure~\ref{fig:Niplume-aitoff} we showcase Aitoff projections of the NiCoFe mass distribution at 19.74$\,$days post-bounce (i.e. at the end of the simulation with $\texttt{P-HotB}$), d$M$/d$\Omega$, which is computed as follows:}
\begin{equation}
    \dfrac{\mathrm{d}M}{\mathrm{d}\Omega} = \int_0^R \rho\, X_\mathrm{NiCoFe}\,r^2\,\mathrm{d}r,
    \label{eq:dMdOmega}
\end{equation}
\txtred{where $\rho$ is the matter density, $X_\mathrm{NiCoFe}$ is the mass fraction of $^{56}$Ni and its decay products $^{56}$Co, $^{56}$Fe, and $R = R(M_\mathrm{ej})$ is the radius corresponding to a specific mass coordinate $M_\mathrm{ej}$ in the ejecta (i.e., the chosen value $M_\mathrm{ej}$ of the ejecta mass is enclosed by radius $R(M_\mathrm{ej})$). The three panels show the Aitoff projections for $M_\mathrm{ej} = 0.1\,M_\odot$, $M_\mathrm{ej} = 2.0\,M_\odot$, and for the full ejecta, from top to bottom in Figure~\ref{fig:Niplume-aitoff}. Additionally, in each panel we denote the locations of the primary $^{56}$Ni plume (red cross) and of a secondary plume (orange plus), as well as the directions of the hydrodynamical neutron star kick (light blue plus), the neutrino-induced kick (yellow plus), and the total (hydrodynamical and neutrino-induced) neutron star kick (gray cross). The Aitoff projections are shown in the coordinate frame after the rotation mentioned above.}

\subsection{Comparison to \J18}
\txtred{Throughout this work, at several points we will compare the newly obtained 3D results from s9.0 to the 1D results from \J18, who modelled a 1D explosion of the same progenitor star in the nebular phase with \texttt{SUMO}. However, some relevant differences between their 1D model and our 3D model are outlined here.

In \J18, the explosion energy is $0.11\,\times10^{51}\,$erg, while here our explosion energy is about a factor two lower at $0.054\times10^{51}\,$erg. As the total ejecta mass is practically the same ($7.4\,M_\odot$ in \J18 to $7.32\,M_\odot$ here), this leads to lower bulk expansion velocities in the 3D model, and higher densities at later times compared to \J18. Additionally, the 1D model is strongly stratified, with all of the $^{56}$Ni contained inside the innermost \kms{450}. Conversely, in 3D the radioactive material is strongly present in the innermost \kms{300} and a few narrow, fast-moving plumes up to approximately \kms{1500} (see Figure~\ref{fig:Niplume-aitoff}), and with a minor presence throughout the total ejecta. The total $^{56}$Ni mass in \J18 is $6.2\times10^{-3}\,M_\odot$, while in 3D it is only marginally lower, with $6.06\times10^{-3}\,M_\odot$.

In the 1D model, the H and He-rich envelope had its solar abundance composition restored for N, Na, Al, K, Sc, Ti, V, Cr, Mn, Co and Ni; in the 3D model here we do not make such an adjustment, and therefore features from e.g. Na and Ni will be missing (as we have no stable Ni isotopes in the 3D model). Out of all these species, only N and Na have abundances that exceed $10^{-4}$ (with $2.7\times10^{-4}$ and $10^{-4}$, respectively). Ti, Cr and Co are included here, without solar composition correction.}

%In Figure~\ref{fig:composition_J18}, the start of the H envelope is shown (from \kms{460}; the whole H envelope has the same composition) -- the abundances of N and Na climb over $10^{-4}$ but the other elements (Al, K, Sc, Ti, V, Cr, Mn, Co, stable Ni) have lower mass fractions and are not included in the Figure.%, but are accounted for in both \texttt{SUMO} and \texttt{ExTraSS}.

% \begin{figure}  %% Not sure if I like this as column-width instead of full-page
%     \centering
%     \includegraphics[width=\linewidth]{Figures/TEMP-composition_J18.pdf}
%     \caption{\txtred{A zoom in on the mass fractions of \J18, in velocity space to better highlight the zoning structures. At higher velocities, the composition remains the same as at \kms{500}.}}
%     \label{fig:composition_J18}
% \end{figure}

% \txtred{In \citet{vanbaal2025ExTraSS}, a more detailed comparison between \texttt{ExTraSS} and \texttt{SUMO} is given, alongside a direct comparison of the \J18 model and a 3D version thereof.}

\section{Results} \label{sec:results}
In this section we will first focus on the physical conditions in the ejecta and then \txtred{on} spectral evolution of s9.0 at two different epochs (250\,d and 400\,d), making comparisons to SN~1997D \citep[the first LLIIP;][]{turatto1998peculiar,benetti2001fading}. We will then examine the line profile variations for different viewing angles for the same element, and between different elements, to investigate the impact of the 3D structure of the ejecta. We will study how well the line profiles can match observed line profiles of SN~1997D and SN~2016bkv \citep[which potentially has an ECSN origin;][]{hosseinzadeh2018short}. 

For SN~1997D, the exact epoch is uncertain due to the late detection of this particular SN. The explosion is estimated to be $50-100\,$days prior to detection \citep[see][]{benetti2001fading,zampieri2003peculiar}, with nebular phase spectra taken 250 and 384 days post-detection. We assume the explosion epoch to be $75\,$days before detection, which puts these spectra to be 325 and 459 days post-explosion. SN~1997D has a redshift of $z=0.004059$, which gives a recession velocity of \kms{1217}. SN~2016bkv was detected within $\sim3\,$days post-explosion \citep{hosseinzadeh2018short}, placing the nebular phase spectra at 259 and 438 days, respectively. The redshift for this object is $z=0.002$, giving it a recession velocity of \kms{600}. The redshifts were obtained from \href{https://www.wiserep.org}{WiSeREP} \citep{wiserep2012}. No large extinction has been reported for either SN, and we did not implement any correction for extinction.

\begin{figure}
    \centering
    \includegraphics[width=\linewidth]{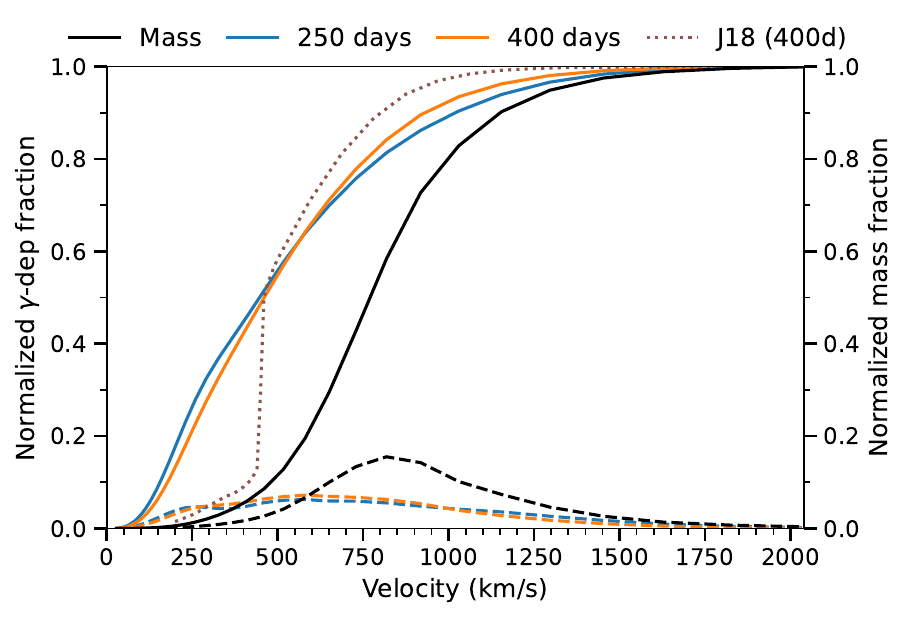}
    \caption{The cumulative $\gamma$-ray energy deposition (coloured lines) and cumulative mass (black). Solid lines show the cumulative fraction, while dashed lines show the fraction per unit velocity. The 400 day deposition curve in the 1D model of \J18 is shown in dotted brown: the dense ``wall'' present in 1D gives a sharp rise at \kms{\sim400} which does not happen in 3D. Instead, there is a distributed deposition in the \kms{100-400} range (receiving around $40\,\%$ of the total power).}
    \label{fig:gamma_evol}
\end{figure}

\subsection{Physical conditions}
\begin{figure}
    \centering
    \includegraphics[width=\linewidth]{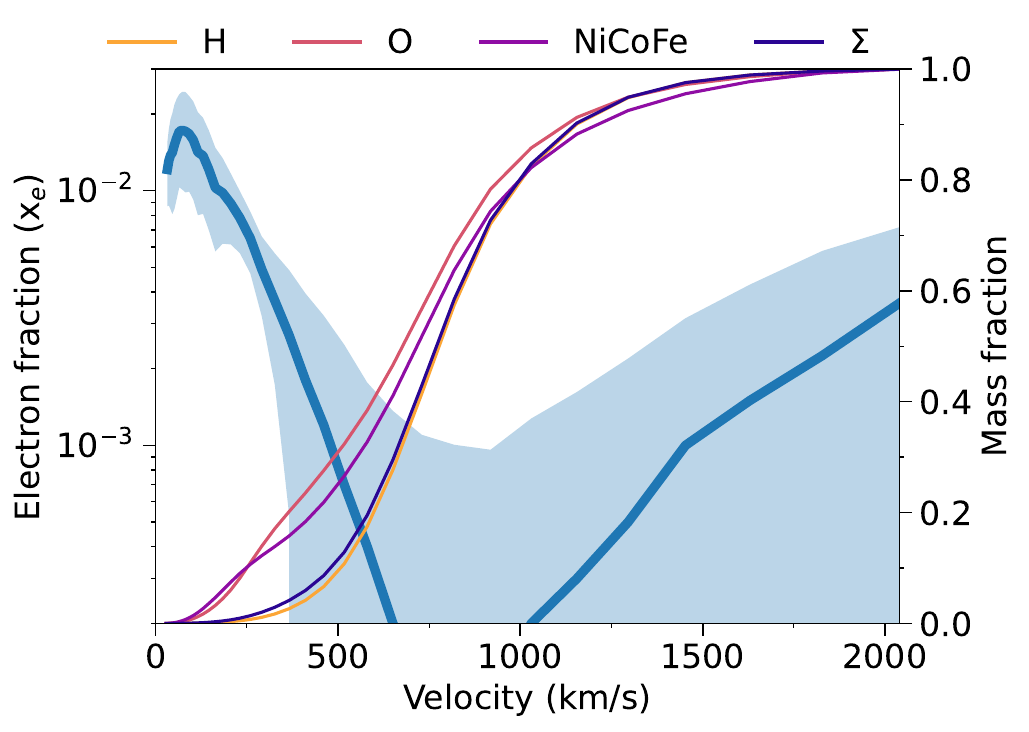}
    \includegraphics[width=\linewidth]{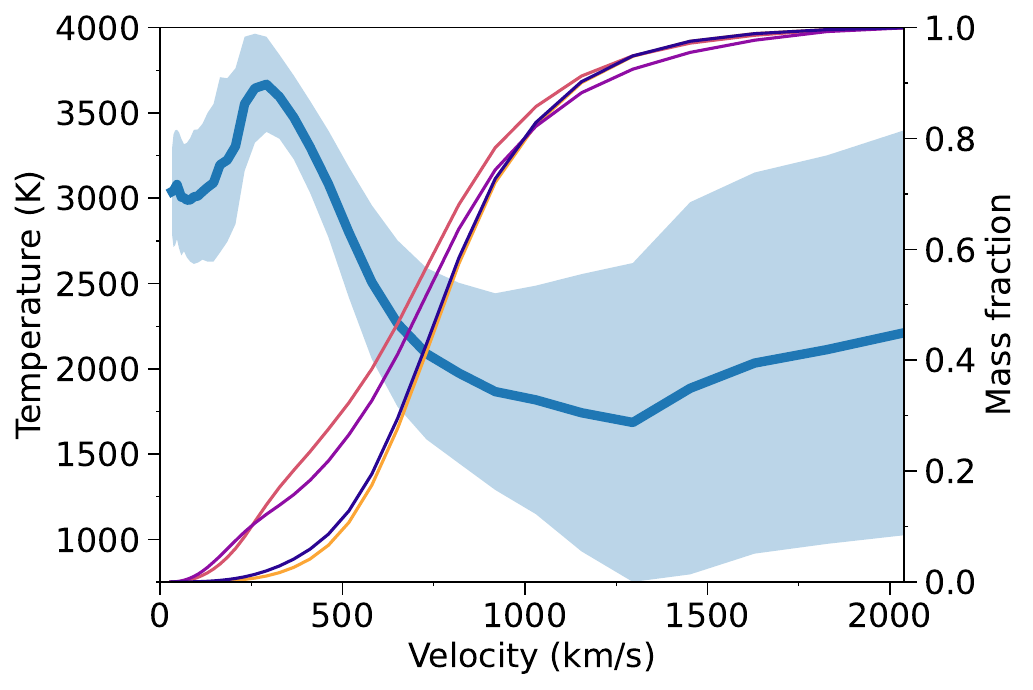}
    \caption{Electron fraction ($x_e$, top) and temperature (bottom) in the s9.0 model at 400 days. Angle-averaged values are plotted as dark blue lines, and $1\sigma$ angle variation as light blue shaded region. Also shown are cumulative distributions of H, O, \txtred{NiCoFe}, and total ejecta.}
    \label{fig:T-X_e}
\end{figure}
\begin{figure*}
    \centering
    \includegraphics[width=\linewidth]{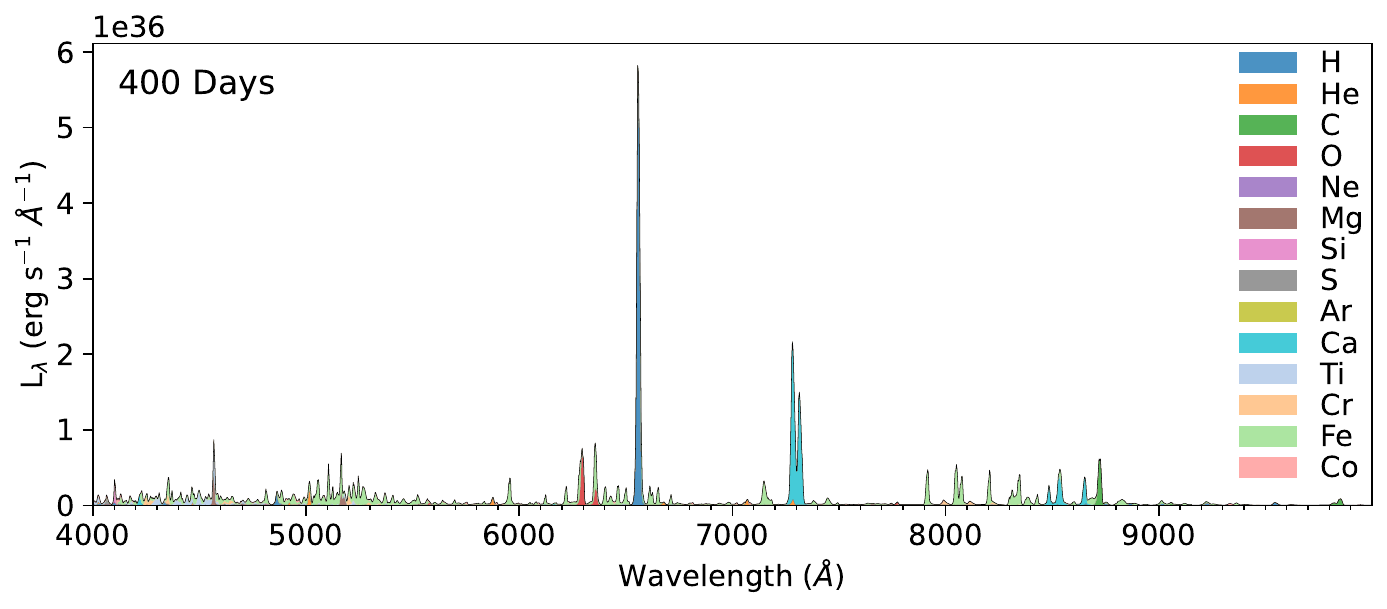}
    \caption{The spectrum of s9.0 at $400\,$days, colour coded by emission origin, in the wavelength range $4000 - 10000\,\angstrom$, for the observer that is most directly approached by the neutron star. The spectrum has a resolution of R=2665.}
    \label{fig:optical_spectrum}
\end{figure*}
In Figure~\ref{fig:gamma_evol}, the cumulative deposition of $\gamma$-rays throughout the ejecta is shown at the two investigated epochs of 250 (blue) and 400 (orange) days, alongside the cumulative mass (in black). What can be seen is that unlike in \txtred{\J18 (whose 400 day $\gamma$-ray deposition is plotted as brown dotted line for comparison)}, deposition occurs throughout the entire ejecta relatively smoothly, as there is no artificial shell structure of O/He layers which absorbed a big part of the $\gamma$-rays in the 1D model. The total energy deposition at 400 days is only at \txtred{$82\,\%$} of what was deposited in the 1D model ($2.37\times10^{39}\,$erg versus \txtred{$1.94\times10^{39}\,$}erg), predominantly due to a lower trapping rate in 3D -- the $^{56}$Ni mass was $6.2\times10^{-3}$ in the 1D model, and is $6.06\times10^{-3}$ here (only 3\% lower). The dashed lines indicate that the energy deposition happens more centrally than where the most mass is present, as the radioactive source is located mostly towards inner region \txtred{(but with noticeable Ni-rich plumes extending far into the envelope, see Figure~\ref{fig:Niplume-aitoff})}. 

In Figure~\ref{fig:T-X_e} the angle-averaged free electron fraction ($x_e$, top) and temperature (bottom) are displayed in blue together with a shaded $1\sigma$ region throughout the ejecta at 400 days. Also shown are the H, O, \txtred{NiCoFe} and total ejecta ($\Sigma$) curves. For the temperature, it can be noted that for the innermost, iron-group rich ejecta the temperature varies around $3000-3600\,$K, being relatively hot as these regions trap a lot of the $\gamma$-rays. At higher velocities, the temperature drops, reaching temperatures around \txtred{$1800\,$}K or lower for the least dense regions, as deposition in these becomes very low (see Figure~\ref{fig:gamma_evol}). \txtred{Past \kms{500}, a significant fraction of the model drops to temperatures below $2000\,$K, which is cold enough that dust and/or molecules could begin to form. However, neither of these are accounted for in \texttt{ExTraSS} in the current version.} The $x_e$ curve is markedly different, with only a very small increase in the innermost ejecta to $\sim0.015$ before dropping by more than an order of magnitude for most of the ejecta, and only increasing again in the outer regions due to recombination becoming less efficient at lower densities. \txtred{In the densest part of the ejecta, between \kms{700\text{ and }1000}, $x_e$ drops below $10^{-4}$.}

Compared to the temperatures and $x_e$ values in \J18 (their Figures~5 and 6), our temperature structure here is more uniform. Their 1D model transitions rapidly from metal-rich core material to envelope composition, which at the inner edge then absorbs a lot of $\gamma$-rays leading to a big temperature increase. Here, such a temperature increase also occurs but for the mixed iron-group rich ejecta itself, which are more efficient coolants and thus drive down the temperatures throughout the ejecta. The lack of stratification in the 3D model also plays a role in the lower $\gamma$-ray trapping.

Our $x_e$ values are similar to the values in the He and H zones in the 1D model -- the relatively high ionization in the pure Fe and O zones in 1D does not occur anywhere in 3D. Another contributing reason for this is that the core densities are higher in the 3D model, as can be seen in Figure~\ref{fig:composition_stock}. Our ejecta up to \txtred{\kms{450} contain nearly $0.5\,M_\odot$}, while in \J18 the mass contained within that velocity coordinate was only $\sim0.2\,M_\odot$. With densities higher in 3D, recombination is more effective and thus the gas is more neutral. Also, the outward mixing of the \txtred{NiCoFe}-rich ejecta means that the $\gamma$-ray energy deposition is less concentrated which lowers non-thermal ionization rates.

\subsection{Spectra of s9.0} % Timeseries if there are multiple!
In Figure~\ref{fig:optical_spectrum} the spectrum of s9.0 at 400 days is shown, colour coded by element for the optical to NIR wavelength range. The strongest lines which appear are H$\alpha$ and the [Ca~II] $\lambda\lambda\,7291,\,7323$ doublet, followed by many lines of similar strength including Mg~I] $\lambda\,4571$, [O~I] $\lambda\lambda\,6300,\,6364$ \txtred{(including a contaminating Fe~I feature on the $6364\,\angstrom$ component)}, a series of Fe~I lines from $7900$--$8400\,\angstrom$\txtred{, the Ca~II NIR triplet, as} well as [C~I] $\lambda\,8727$. There are several Fe~I lines also present between $6200$--$6700\,\angstrom$ and at $5957\,\angstrom$. All of these are stronger than, or comparable to, [Fe~II] $\lambda\,7155$ which is the strongest line from the ionized species and one of the most isolated Fe lines in the optical.

\subsubsection{Comparison to SN~1997D}
\begin{figure*}
    \centering
    \includegraphics[width=\linewidth]{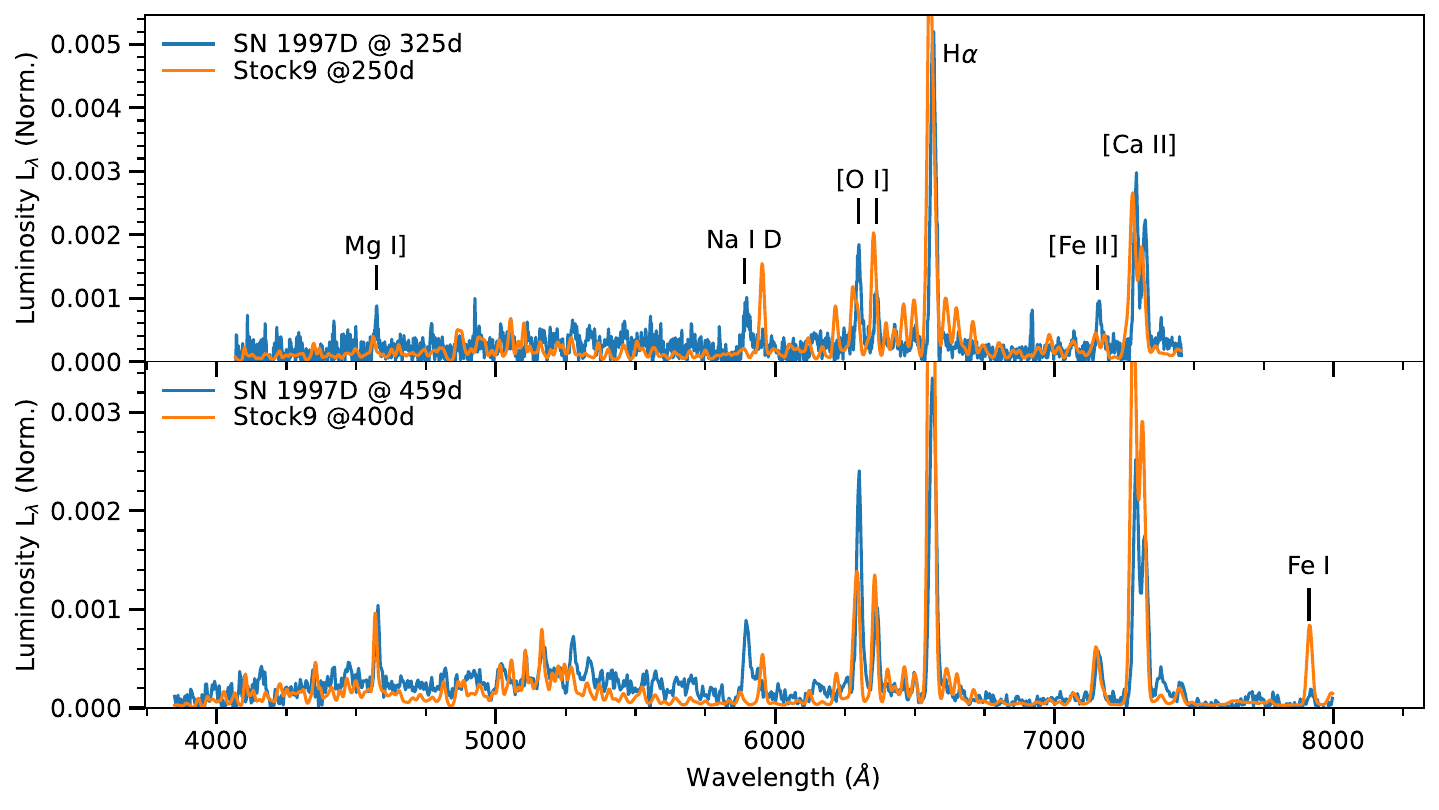}
    \caption{The model spectra for the same viewing angle as in Figure~\ref{fig:optical_spectrum}, in orange, at 250 days (top) and 400 days (bottom) compared to SN~1997D, in blue, at 325 days (top) and 459 days (bottom). All spectra are normalized to the sum of the flux over the observed wavelength range, and are zoomed in to focus on the weaker lines, as the models have significantly stronger H$\alpha$ peaks (the models peak at \txtred{0.0068 and 0.010 for the top and bottom panels, respectively}). The model spectra are convolved to the spectrograph resolutions ($R=579$ and $R=457$, respectively).}
    \label{fig:PIB-SN1997D}
\end{figure*}
\begin{figure*}
    \centering
    \includegraphics[width=\linewidth]{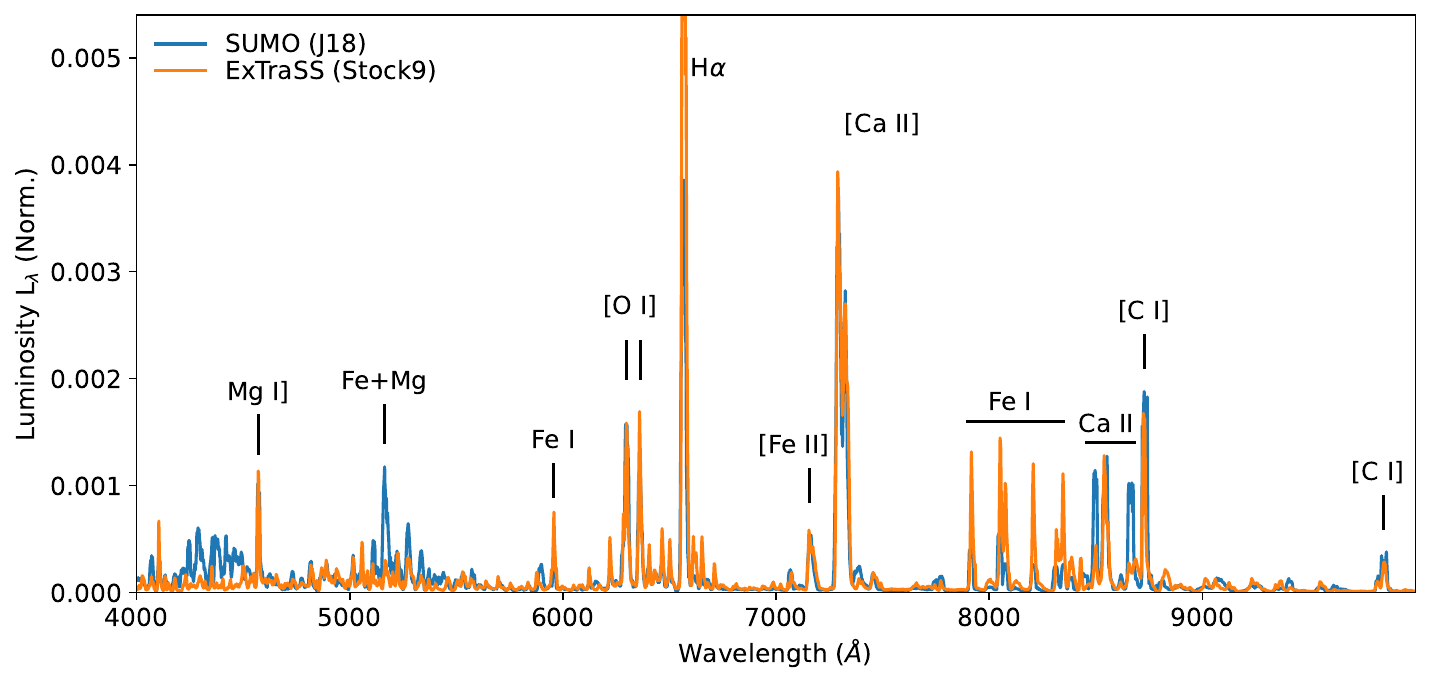}
    \caption{Comparison of \txtred{normalized} model spectra (same viewing angle as in Figure~\ref{fig:optical_spectrum}) in the optical range for both \txtred{\texttt{SUMO} (blue) from \J18, and \texttt{ExTraSS} (orange), both} at 400 days. The H$\alpha$ peaks in the \texttt{ExTraSS} model at \txtred{0.011}.}
    \label{fig:SUMO-ExTraSS_spectral}
\end{figure*}
In Figure~\ref{fig:PIB-SN1997D} we show the model spectra from \txtred{s9.0} compared against SN~1997D at two epochs, at 250/325 days (model/observation) and at 400/459 days. The chosen model viewing angle is the same as in Figure~\ref{fig:optical_spectrum}, i.e. the viewing angle most directly aligned with the neutron star motion. The strongest observed features are marked in the top panel, alongside Fe~I $\lambda\,7912$ in the bottom panel. Aside from H$\alpha$, which is much stronger in the model than the other lines, the line strengths match reasonably well at both epochs. Mg~I] $\lambda\,4571$ is on the weak side at 250 days but very close at 400 days, while [Ca~II] $\lambda\lambda\,7291,\,7323$ is $\sim30\,\%$ stronger in the model than the observations at \txtred{the later} epoch \txtred{and matches well at the first one}. [Fe~II] \txtred{is somewhat weak in} the early epoch, but is a \txtred{good fit} at late times. The [O~I] doublet is a little bit weaker in the model spectra than the observations for the $\lambda\,6300$ feature but a \txtred{decent} match on the $\lambda\,6364$ feature, which is due to the relatively strong Fe~I contaminating feature in the models. The observed spectra show a strong Na~I $\lambda\lambda\,5890,\,5896$ line (Na~I D in Figure~\ref{fig:PIB-SN1997D}), which is not included in the model.

A key property of the model spectra is the presence of strong Fe~I lines between $7900$--$8400\,\angstrom$. SN~1997D does not have observations in the nebular phase which cover this wavelength range. Other observations which do cover this range include SN~2016bkv, which also has a good timing match to the models here (at 259 and 438 days). While unfortunately these spectra have a strong blue contamination which, without proper removal, prohibits a normalized flux comparison, the strong Fe~I lines in the model spectra here are not observed in SN~2016bkv, giving consistency with the conclusions from the bluer wavelengths of SN~1997D that the model tends to have too strong Fe~I emission.

\subsubsection{Comparison to 1D model of \J18} \label{ssec:J18_1D3D}
\txtred{In Figure~\ref{fig:SUMO-ExTraSS_spectral}, we compare our spectral output from s9.0 (for the same viewing angle as in Figure~\ref{fig:optical_spectrum}) to the 1D \texttt{SUMO} output of \J18. Compared to \J18's 1D model, a few notable differences appear.} H$\alpha$ is significantly stronger, while \txtred{Mg~I], [O~I], [Ca~II] and Ca~II are of similar strength.} The ``H-zone'' spectrum in \J18 (their Figure~9) has similar line ratios between [O~I], H$\alpha$, [Ca~II] and Ca~II NIR to what can be seen in Figure~\ref{fig:optical_spectrum} here. Fe~I emission is significantly stronger in the 3D model in the wavelength ranges $6200$--$6700\,\angstrom$ \txtred{(including the contamination on [O~I]'s $6364\,\angstrom$ component)} and $7900$--$8400\,\angstrom$, but is weaker on the blue ($\leqslant5500\,\angstrom$) side \txtred{-- including the blended Fe+Mg feature at $5180\,\angstrom$}. As the 3D model is less ionized, the emission from Fe~II is also weaker, although [Fe~II] $\lambda\,7155$ is still \txtred{matching well}. The 1D model had a strong [C~I] $\lambda\,8727$ line, which is still present in 3D but is \txtred{somewhat less} prominent. Another feature which the 1D model predicted to be appearing only in a ``pure envelope ejecta'' (a mimic for ECSNe) is O~I $\lambda\,8446$, which is insignificant in the model spectra here. However this line is formed by Ly$\beta$ line-overlapping pumping which is treated by \texttt{SUMO} but not by \texttt{ExTraSS}. As sodium \txtred{(Na)} is not included in the nucleosynthesis in \texttt{Prometheus}, and we made no correction towards solar abundance patterns for missing elements (which was done in the 1D model), the Na~I D line is missing compared to \J18 \txtred{-- the nearby emerging feature in \texttt{ExTraSS} is He~I $5876\,\angstrom$}. \txtred{The strengthening of the Fe~I features in 3D} compared to 1D can be attributed to the higher degree of mixing and stronger presence of synthesized iron in larger regions of the ejecta.

There were several features in \J18 which were attributed specifically to the hydrostatic nucleosynthesis layers of the He core and considered signatures of this, distinguishing a Fe CCSN event from an ECSN one. These were He~I $\lambda\,7065$, [C~I] $\lambda\,8727$, [C~I] $\lambda\lambda\,9824,\,9850$, and O~I $\lambda\,8446$. In addition, the He core material strengthened Mg~I] $\lambda\,4571$ and [O~I] $\lambda\lambda\,6300,\,6364$. With these layers experiencing mixing in 3D, a key question is whether this conclusion based on 1D analysis still holds. Figure~\ref{fig:SUMO-ExTraSS_spectral} shows that it does; these features remain present in the 3D model and at roughly similar luminosities. While the He~I $\lambda\,7065$ and O~I $\lambda\,7774$ are quite weak both in 1D and 3D, and can be hard to detect in practice, the other lines are easily detectable according to both the 1D and 3D models. The O~I $\lambda\,8446$ line \txtred{has no special treatment in \texttt{ExTraSS} (see above) and thus} is of negligible strength here.

\subsection{Line profile properties}
\begin{figure*}
    \centering
    \includegraphics[width=.49\linewidth]{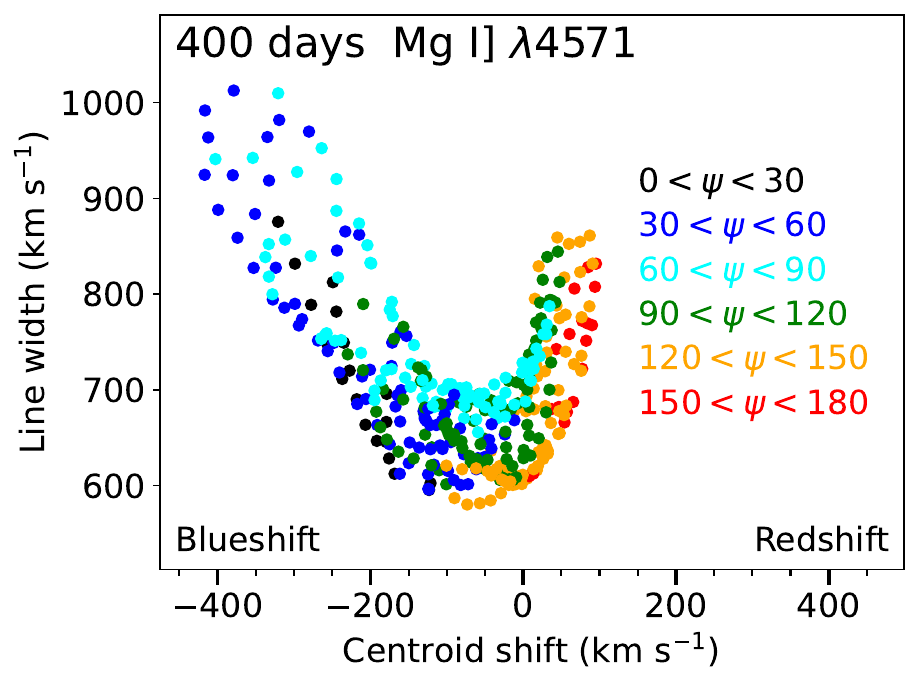}
    \includegraphics[width=.49\linewidth]{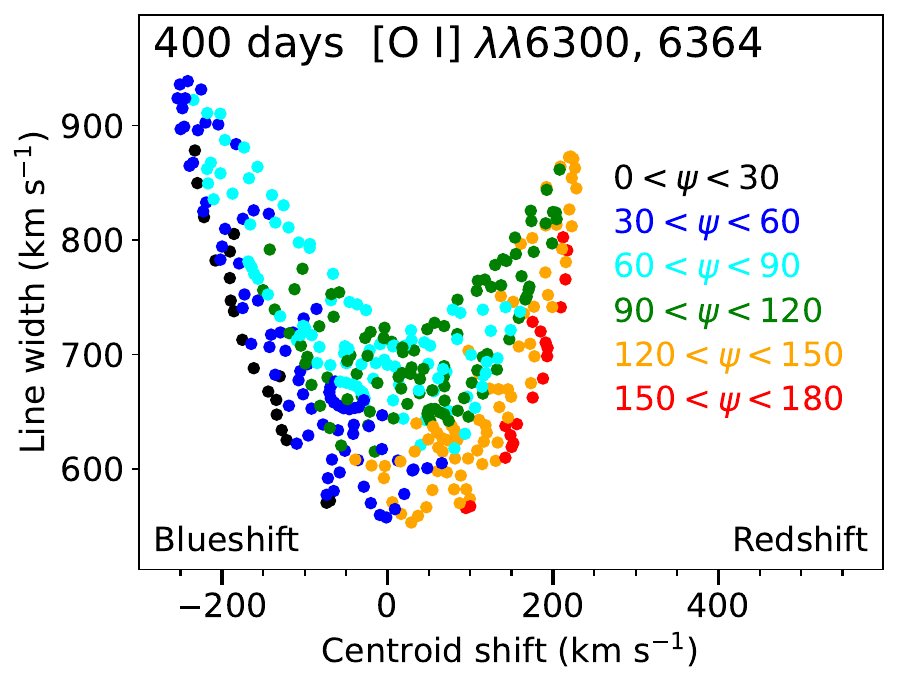}
    \includegraphics[width=.49\linewidth]{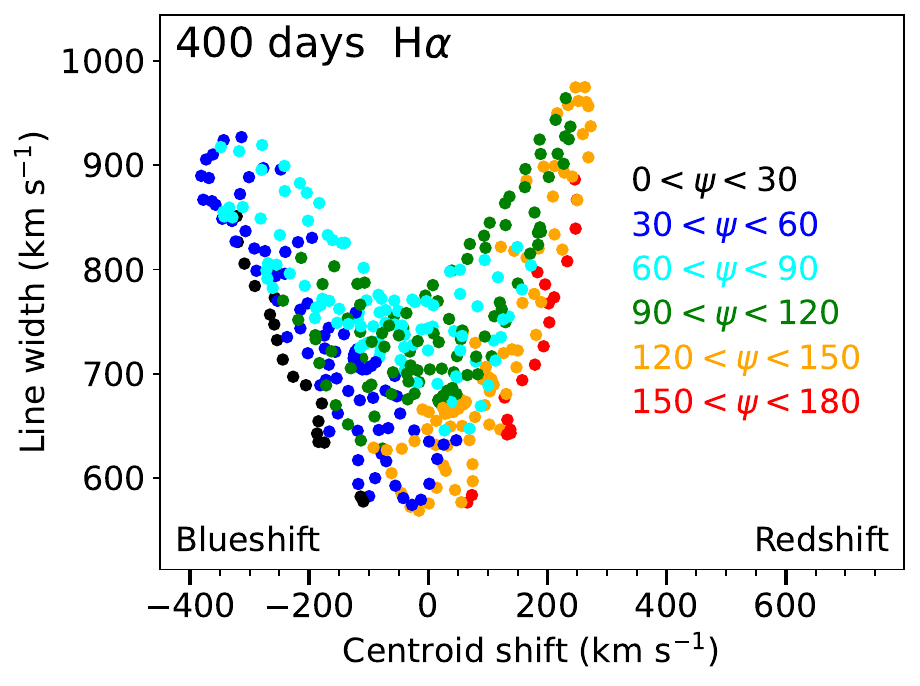}
    \includegraphics[width=.49\linewidth]{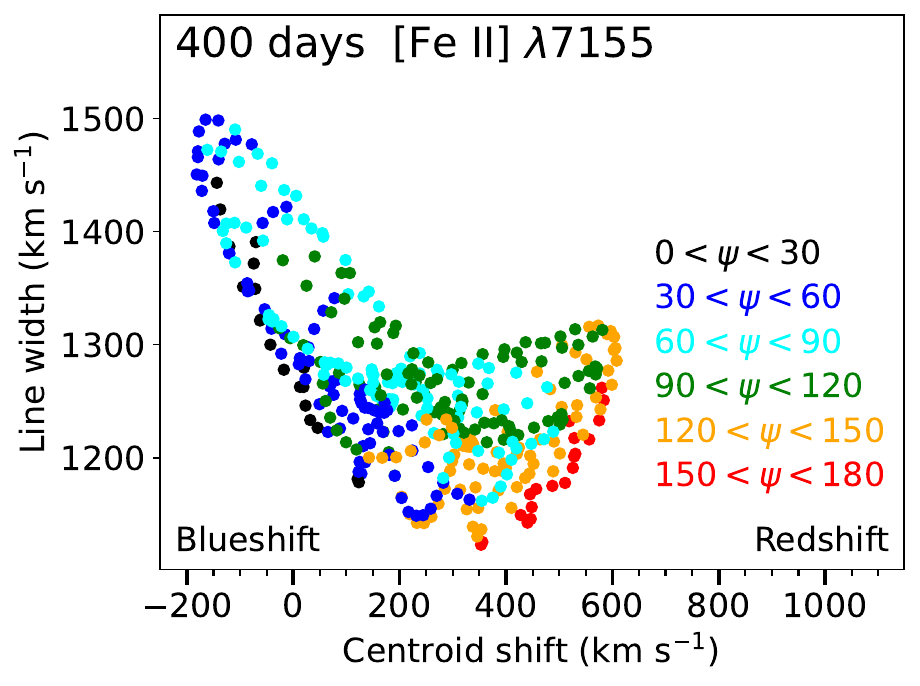}
    \includegraphics[width=.49\linewidth]{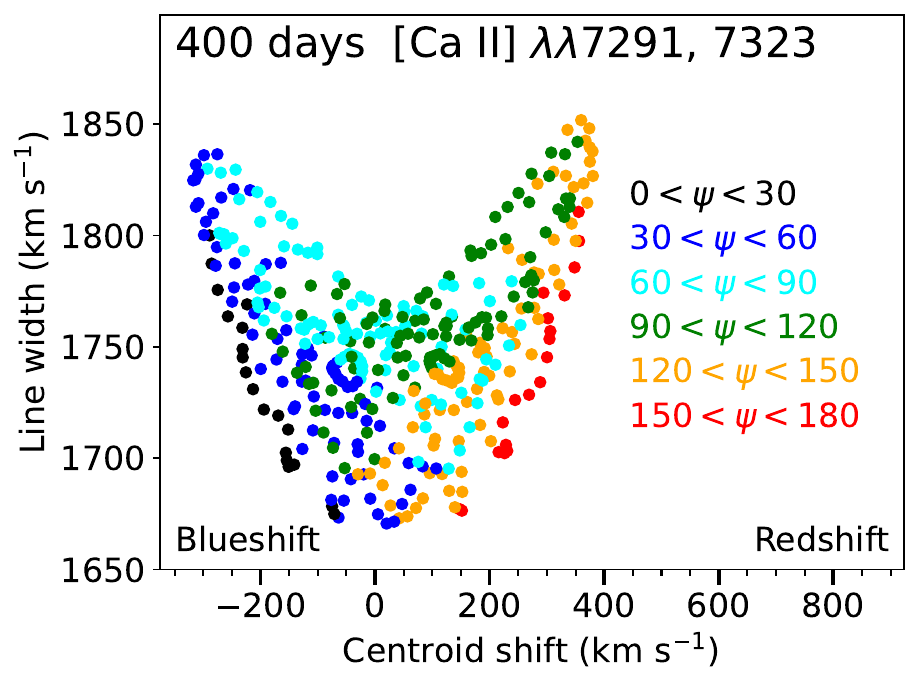}
    \includegraphics[width=.49\linewidth]{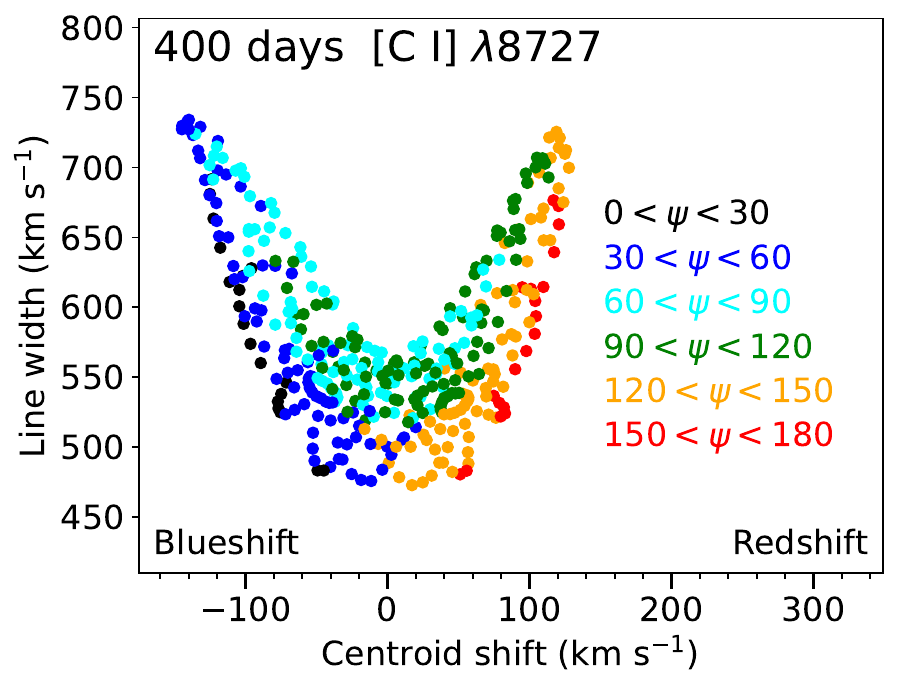}
    \caption{The line properties for the six spectral features Mg~I] $\lambda\,4571$ (top left), [O~I] $\lambda,6300$ (top right), H$\alpha$ (mid left), [Fe~II] $\lambda\,7155$ (mid right), [Ca~II] $\lambda\lambda\,7291,\,7323$ (bottom left) and [C~I] $\lambda\,8727$ (bottom right) for all viewing angles, colour coded for the angle $\Psi$ which is the angle between the direction vector to the viewer and the neutron star motion vector, i.e. the black points (small $\Psi$) correspond to viewing angles where the neutron star is moving almost directly towards the observer. Centroid shifts are given relative to the rest wavelength of the singlet features and [O~I] $\lambda6300$, and relative to the transition strength weighted rest wavelength for [Ca~II] (at $7304\,\angstrom$). Positive (negative) centroid shifts correspond to a redshifted (blueshifted) centroid. [O~I] is treated as singlet line as the two features are not blended together due to the low velocity.} 
    \label{fig:profile_props}
\end{figure*}
In Figure~\ref{fig:profile_props}, the line properties (centroid shift and line width\footnote{See Eqs. 2 and 3 in \citet{vanbaal2023modelling} for the definitions. $v_\text{shift}$ corresponds to the line centroid shift, while $v_\text{width}$ corresponds to the feature's width. In \citet{jerkstrand2020properties} these are also defined in Eqs. 7 and 8.}) are shown for six of the strongest features: Mg~I] $\lambda\,4571$, [O~I] $\lambda\,6300$ (the [O~I] properties are calculated from purely the $6300\,\angstrom$ feature as the doublet is resolved), H$\alpha$ $\lambda\,6563$, [Fe~II] $\lambda\,7155$, [Ca~II] $\lambda\lambda\,7291,\,7323$ and [C~I] $\lambda\,8727$. For all features, only the emission from the specific element is used (so e.g. [O~I] $\lambda\,6300$ is not influenced by the nearby Fe~I feature at $6280\,\angstrom$). \txtred{The viewing angles are colour coded for the angle $\Psi$, which is the angle between the direction vector to the viewer and the neutron star motion vector (the gray cross in Figure~\ref{fig:Niplume-aitoff}), i.e. small $\Psi$ (black points) corresponds to viewing angles where the neutron star is moving almost directly towards the observer.}

The patterns for these six lines are relatively similar, although certain differences appear. Most of the lines have a width of at least \kms{500-600} and reach up to \kms{\sim900-1000} for some viewing angles. [C~I] has the lowest peak widths at \kms{\sim\txtred{750}}, despite being the lightest emitting element of these six aside from hydrogen (and would have higher velocities than O, Ca, Fe in 1D). It \txtred{also has the lowest} minimum width (of the \txtred{unblended} singlet lines) at \kms{\txtred{\sim500}}. While the amount of synthesized carbon is not very large, it is still sufficient to give distinct line emission, stronger than that of the primordial carbon. The viewing angle variation of the C lines therefore indicates a less asymmetric distribution than for heavier elements, in line with the carbon residing far out in the He core at explosion. 

[Ca~II] $\lambda\lambda\,7291,\,7323$ is markedly wider than the other lines due to the doublet nature which even at these low expansion velocities is still blended together (while [O~I] is resolved into two distinct peaks). [Fe~II] $\lambda\,7155$ however is a singlet line and still comes out as much wider than the others due to contamination of the nearby [Fe~II] $\lambda\,7172$ (see also \citealt{jerkstrand2015supersolar}, where the luminosity of this line is around $24\%$ of [Fe~II] $\lambda\,7155$). Mg~I] and [Fe~II] have their widest profiles on the blueshifted side, while [O~I], H$\alpha$, [Ca~II] and [C~I] do not show such a preference. Compared to \J18, the profiles come out a bit narrower or just matching the 1D widths, which were \kms{\sim900} for the He core lines and \kms{\sim1100} for H$\alpha$.

For the centroid shifts, H$\alpha$ has the biggest spread with the most blueshifted centroids reaching \kms{-400} and the most redshifted ones \txtred{\kms{300}}. [C~I] instead has the smallest spread, with the centroid shift ranging from \kms{-150\text{ to }150}. Both [Fe~II] and [Ca~II] also have mostly redshifted centroids, with values ranging from \kms{\txtred{-200}\text{ to }600} and \kms{\txtred{-300}\text{ to }400}, respectively. Mg~I] is quite irregular, unlike H$\alpha$, and has smaller centroid shifts which range between \kms{\txtred{-400}\text{ to }\txtred{100}}. The [O~I] profile is very similar to H$\alpha$ and [C~I], but with intermediate centroid shifts from \kms{\txtred{-250}\text{ to }250}.

Generally, for all elements the trend is that viewers with a smaller $\Psi$ (i.e. the neutron star is more directly approaching the viewer) see the most strongly blueshifted line profiles, while the viewers with large $\Psi$ see the most redshifted profiles. In explosions with a large hydrodynamical kick, spectral lines of elements between Mg and Fe should exhibit a tendency of being redshifted when $\Psi$ is small, as these elements are preferentially produced and ejected by the stronger SN explosion in the direction opposite to the neutron star kick vector \citep{wongwathanarat2013three}. Oxygen might show the opposite behaviour, i.e. blueshifted lines for small $\Psi$, if it is only present in a small amount in the progenitor, as the stronger SN explosion opposite to the neutron star kick vector would lead to more explosive oxygen burning.

In the model here, however, the neutron star kick is dominated by the \txtred{neutrino}-induced kick instead of the hydrodynamical kick \citep{janka2024interplay}, and therefore the correlations described above are not expected to hold. The angle between the two kick vectors is $82.4^\circ$ \txtred{(see Figure~\ref{fig:Niplume-aitoff})}, with a total kick velocity of \kms{v^\text{tot}_\text{NS}=57}. As this kick velocity is quite low, secondary NLTE effects play an important role in how the centroid shifts align with $\Psi$. When we perform the calculation against $\Psi_\text{hydro}$, accounting only for the hydrodynamical kick, the colour patterns become very irregular. This indicates that there is no \hbox{(anti-)}alignment towards the hydrodynamical kick alone, which is in agreement with expectations for low kick velocities. \txtred{A lack of} $\Psi:v_\text{shift}$ aligning effect was also seen in the model of \citet{vanbaal2023modelling}, which similarly had a relatively weak neutron star kick. \txtred{In Section~\ref{ssec:NiPlume} we take a more detailed look at the impacts of the strong Ni plume which is present in the model (see also Figure~\ref{fig:Niplume-aitoff}), both for the line profile properties and the line profiles themselves.}
%There is still some variations between the different elements, as in particular [Fe~II] comes out to favour redshifted centroids, but this is due to a weak contaminating line on that side.

\begin{figure*} %\txtred{(If I want to normalize to $\Sigma$flux, need to invert row-column)}
    \centering
    \includegraphics[width=\linewidth]{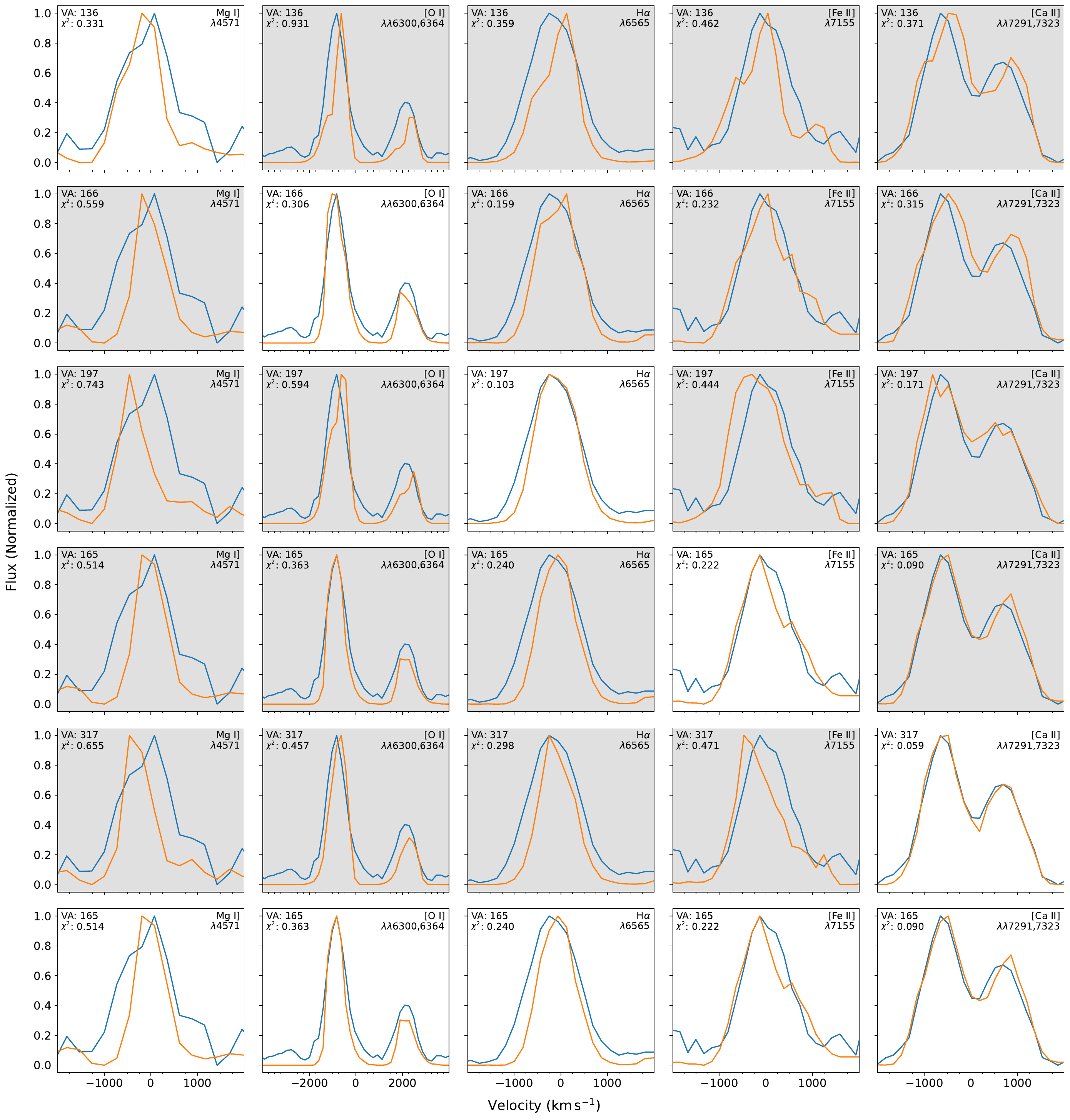}
    \caption{A comparison between SN~1997D at 459 days (blue) to the model spectra at $400\,$days (orange). For rows 1-5, the viewing angle is the one giving the best $\chi^2$ fit for the line with white background. In column order; Mg~I] $\lambda\,4571$, [O~I] $\lambda\lambda\,6300,\,6364$, H$\alpha$, [Fe~II] $\lambda\,7155$, [Ca~II] $\lambda\lambda\,7291,\,7323$. For row six, the viewing angle is the one giving best overall fit for all lines together. The centres of the [O~I] profiles are at $6316\,\angstrom$, and at $7304\,\angstrom$ for [Ca~II].}
    \label{fig:lineprofdet_sn1997D}
\end{figure*}
\subsection{Line profile details}
It is also important to check if (some) of the synthetic line profiles are good matches to observed spectral lines, as this indicates if the explosion models are generating a realistic ejecta morphology or not. In this section, the line profile variations will first be investigated against observations from SN~1997D and SN~2016bkv, and then also against each other.

\subsubsection{Comparison to SN~1997D}
In Figure~\ref{fig:lineprofdet_sn1997D}, a comparison between the line profiles of Mg~I], [O~I], H$\alpha$, [Fe~II] and [Ca~II] from the model spectra (orange) are shown versus the observations of SN~1997D at 459 days (blue). The model spectra are convolved to the same resolution as the observations ($R=457$, or \kms{650}). Each profile is \txtred{separately} normalized to its peak value. \txtred{As Figure~\ref{fig:PIB-SN1997D} shows, the line profile ratios in the model are not quite the same as in SN~1997D, preventing a singular normalization to the whole profile.} Each element has its own column; each row is a separate viewing angle. The rows are arranged by the viewing angle that best fits 1) Mg~I], 2) [O~I], 3) H$\alpha$, 4) [Fe~II], 5) [Ca~II] and lastly 6) fitting all five profiles together; the ``best fit'' is decided through a $\chi^2$ fit between the observed spectrum and interpolated model. \txtred{The $\chi^2$ values are normalized to the highest $\chi^2$ value for that element across all the viewing angles, meaning that these $\chi^2$ values should not be compared between different elements or datasets of different SNe.} For the oxygen profile, only the oxygen emission is used, as observationally there is no clear contamination while the models have strong overlapping Fe~I emission. For all other lines, the full emission profile in the model is used.

For Mg~I], it can be seen that the best-fitting viewing angle (\txtred{VA 136}) can match the observed profile quite well, although the observed profile has a small bump at \kms{\sim1000} which is not reproduced for (most) viewing angles. The viewing angles which are fitting well for the other elements typically have narrower Mg~I] profiles (see the other rows), leading to poor $\chi^2$ values for these. Conversely, \txtred{VA 136} has somewhat too narrow profiles for [O~I] and H$\alpha$, and to a lesser extent for [Fe~II].

For [O~I] (\txtred{VA 166}), the model spectra use only the oxygen-emitting component and the Fe-contamination is thereby removed (see Figure~\ref{fig:optical_spectrum}); in the observations the red component of this doublet might still contain some Fe emission which might explain why none of the viewing angles have a great match for the [O~I] $\lambda\,6364$ component. The best [O~I] viewing angle also has a decent fit for [Fe~II] and [Ca~II] as well, but has too narrow Mg~I] and H$\alpha$ profiles.

The angle where H$\alpha$ matches best (\txtred{VA 197}) \txtred{does not do too well for any of the other elements. For} for Mg~I] this viewing angle comes out very narrow, and for [Ca~II] the fit is not good as the double-peaked nature of the doublet \txtred{partially} vanishes at this viewing angle. 

The viewing angle with the best fitting [Fe~II] profile (\txtred{VA 165}) \txtred{also has a} narrow Mg~I] profile, and a slightly narrow but otherwise well-matching H$\alpha$ profile. It reproduces the [O~I] profile quite decently \txtred{(it is directly adjacent to the [O~I] best fitting angle)}, while the [Ca~II] profile is \txtred{quite a close match to the observed profile}.

[Ca~II]'s best fitting viewing angle (\txtred{VA 317}) is quite far from each of the other best-fitting viewing angles, and subsequently does not match any of the observed line profiles for the other elements very well. Mg~I]\txtred{, [O~I],} and H$\alpha$ are too narrow, although [Fe~II] is reasonable in width and shape.

When determining the viewing angle which gets the best fit for all five line profiles weighted together, the outcome is \txtred{the} viewing angle (\txtred{VA} 165) \txtred{that was also the best fitting angle for [Fe~II]. This angle also has one of the best fits for [O~I], and captures [Ca~II] very well while H$\alpha$ is decent. Mg~I] is not too well fitted here, but as this is the case for the majority of the viewing angles this angle still comes out as the best one overall.}

\subsubsection{Comparison to SN~2016bkv}
\begin{figure*}
    \centering
    \includegraphics[width=\linewidth]{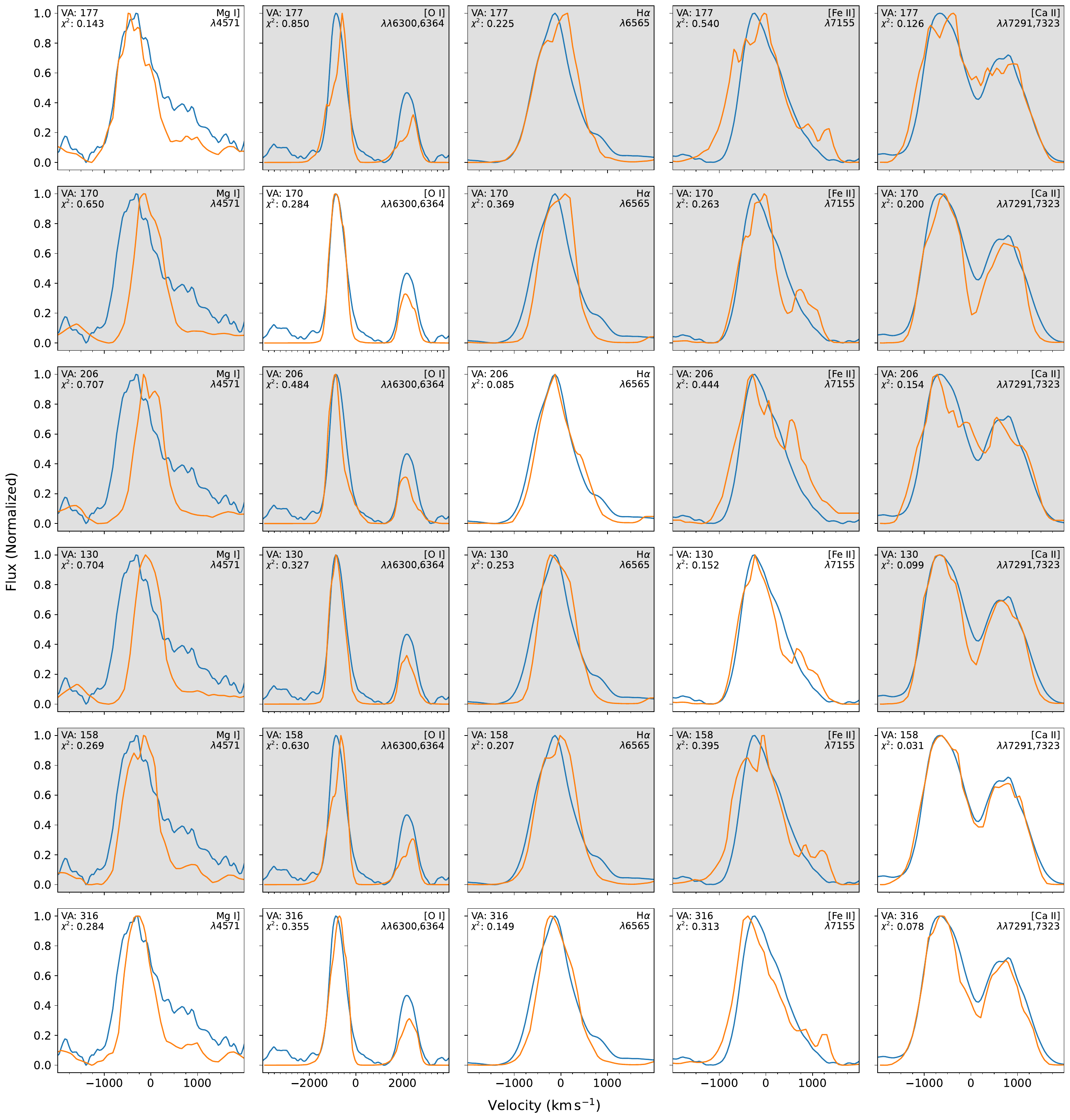}
    \caption{The same as Figure~\ref{fig:lineprofdet_sn1997D}, but against SN~2016bkv (in blue) at 438 days.}
    \label{fig:lineprofdet_sn2016bkv}
\end{figure*}
\begin{table}
 \centering
 \caption{\txtred{$\theta$ and $\phi$ coordinates for the viewing angles (VA) which correspond to at least one of the best-fitting profiles between the model here and SN~1997D at 459 days or SN~2016bkv at 438 days.}}
 \label{tab:VAhelper}
 \setlength\tabcolsep{9pt}
 \begin{tabular*}{.85\linewidth}{c c c c r} 
  \hline
  VA & $\theta$ & $\phi$ & SN & Best feature \\
  \hline
  $\#130$  & $58.5^\circ$ & $171^\circ$ & 16bkv & [Fe~II] \\
  $\#136$  & $58.5^\circ$ & $279^\circ$ & 97D & Mg~I] \\
  $\#158$  & $67.5^\circ$ & $315^\circ$ & 16bkv & [Ca~II] \\
  $\#165$  & $76.5^\circ$ & $81^\circ$ & 97D & [Fe~II] $\&$ Full \\
  $\#166$  & $76.5^\circ$ & $99^\circ$ & 97D & [O~I] \\
  $\#170$  & $76.5^\circ$ & $171^\circ$ & 16bkv & [O~I] \\
  $\#177$  & $76.5^\circ$ & $297^\circ$ & 16bkv & Mg~I] \\
  $\#197$  & $85.5^\circ$ & $297^\circ$ & 97D & H$\alpha$ \\
  $\#206$  & $94.5^\circ$ & $99^\circ$ & 16bkv & H$\alpha$ \\
  $\#316$  & $139.5^\circ$ & $279^\circ$ & 16bkv & Full \\
  $\#317$  & $139.5^\circ$ & $297^\circ$ & 97D & [Ca~II] \\
  \hline
 \end{tabular*}
\end{table}
In Figure~\ref{fig:lineprofdet_sn2016bkv}, the model line profiles (orange) are compared against SN~2016bkv at 438 days (in blue). The resolution of the observation in this epoch is higher than for the model spectra ($R=5000$, or \kms{60}), so the observed spectrum was convolved to the same resolution as our models ($R=2700$, or \kms{111}). \txtred{Also here, the $\chi^2$ values have been normalized by element.}

The viewing angle where the Mg~I] profile fits best (\txtred{VA 177}) creates a very good match to the width of the profile at peak, \txtred{although slightly offset}, while also having a decent match to the bump in the red wing of the observed profile. This bump in the red wing of Mg~I] was also present in the observed spectra of SN~1997D, but is often missing in the model spectra. This viewing angle has, however, poor fits for the other elements\txtred{, in particular for [O~I] and [Fe~II]. Each of the other elements show} a profile which peaks on the red side of the rest wavelength which is not seen in the observations.

For [O~I] (\txtred{VA 170}), the observation might have some Fe contamination in the $6364\,\angstrom$ wing, as in SN~1997D, whereas the model spectra extract only the oxygen-emission, which is likely why every viewing angle underpredicts the strength of the red component. The best fitting viewing angle however does a very good job for the $6300\,\angstrom$ side, \txtred{although the profile fits for the other elements are quite poor relative to their best fits}.

The best angle for H$\alpha$ (\txtred{VA 206}) \txtred{has a bad fit for Mg~I], and is a bit too narrow for [O~I] as well. Additionally, the [Fe~II] and [Ca~II] profiles are quite bumpy at this viewing angle, unlike the observed profiles.} The small offsets around \kms{-700} and \kms{+900} in the observed H$\alpha$ profile could be due to emission from the [N~II] $\lambda\lambda\,6548,\,6583$ doublet, an element not present in our models.

The best-fitting viewing angle for [Fe~II] (\txtred{VA 130}) captures the fast rising blue side and slower declining red side of the profile very well, \txtred{although the red side is a bit bumpy. It has quite a good fit for [Ca~II] as well and the} H$\alpha$ and [O~I] fits are also not bad, but the Mg~I] profile is too narrow here. 

\begin{figure*}
    \centering
    \includegraphics[width=\linewidth]{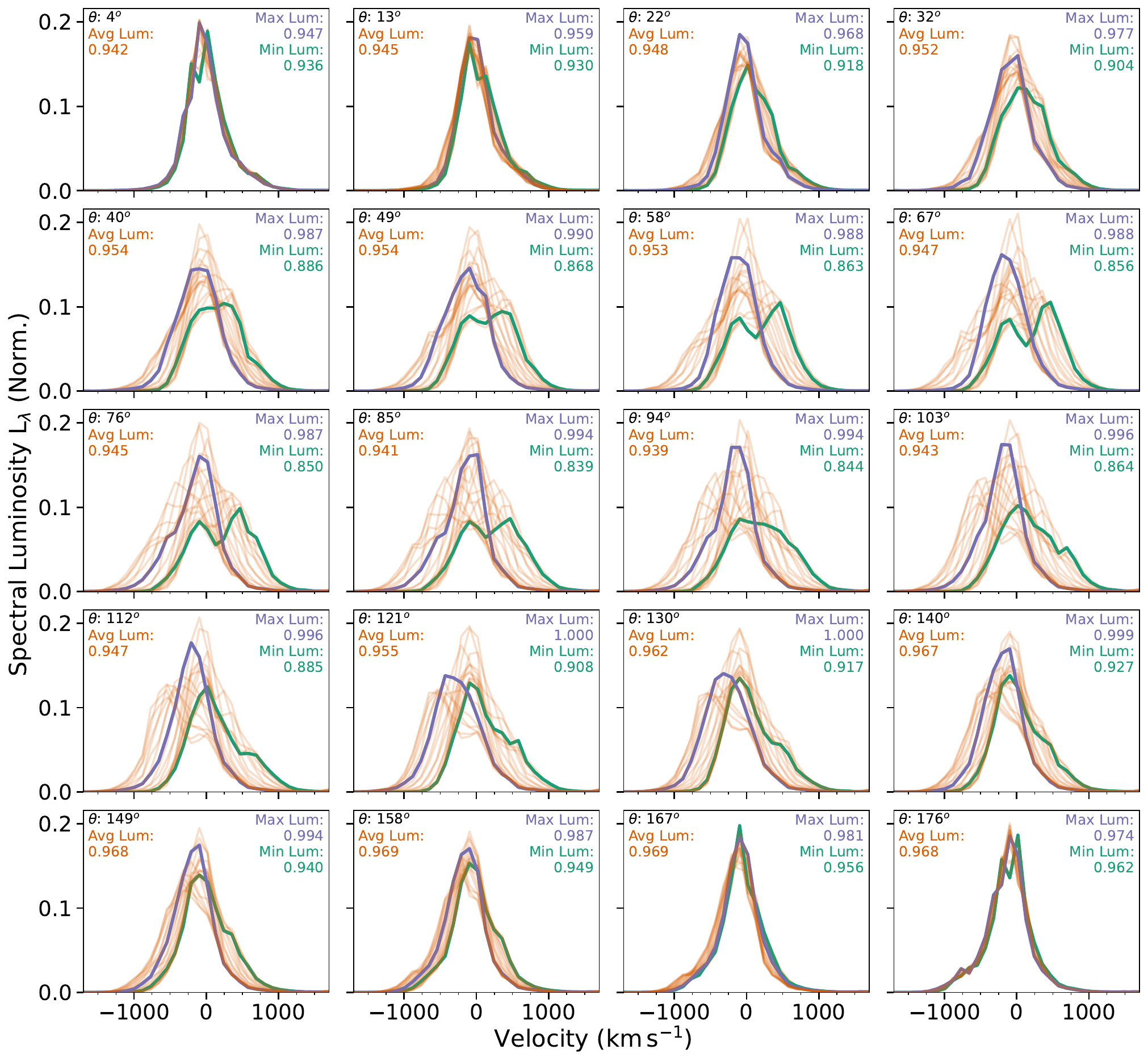}
    \caption{The full set of H$\alpha$-profiles at $400\,$days for all viewing angles, separated by the polar angle $\theta$. In each panel, the set of 20 equatorial observers is shown, with the most luminous line in purple, the least luminous line in green and the others in orange. All line profiles are normalized relative to the most luminous observer. The integrated luminosities for the most and least luminous profiles are noted in the top right of each panel, and the average luminosity in the top left.} % "all line profiles are globally normalized" ?
    \label{fig:Halpha-400_fullVA}
\end{figure*}
The viewing angle with the best [Ca~II] profile (\txtred{VA} 158) gives a very good match. The H$\alpha$ match is also quite good \txtred{and Mg~I] is only slightly too narrow. Both [O~I] and [Fe~II] however} are quite a bit worse \txtred{and all other elements show a peak on the red side of the profile} for this viewing angle, but not in the observations.

When accounting for all elements together, the best viewing angle is \txtred{VA 316}, which is not very close to any of the single-line best viewing angles (which are \txtred{somewhat spread apart for SN~2016bkv but not drastically so}). For \txtred{VA 316}, \txtred{Mg~I] is slightly too narrow and almost completely missing the red wing, while [Fe~II] is somewhat offset to the blue side, but otherwise each of the five elements finds a pretty good fit and thus this viewing angle comes out on top}. 

\txtred{To aid the reader put some perspective on how separated the different viewing angles are, in Table~\ref{tab:VAhelper} we outline the $\theta$ and $\phi$ angles for each of the viewing angles. These are given with respect to the alignment of the model grid, and have no physical bearing towards the neutron star kick or any particular feature of the explosion model. They are merely shown to better highlight how close or far the different observers are that find good line profile fits in the model.}

\subsubsection{Viewing angle variation}
\begin{figure*}
    \centering
    \includegraphics[width=\linewidth]{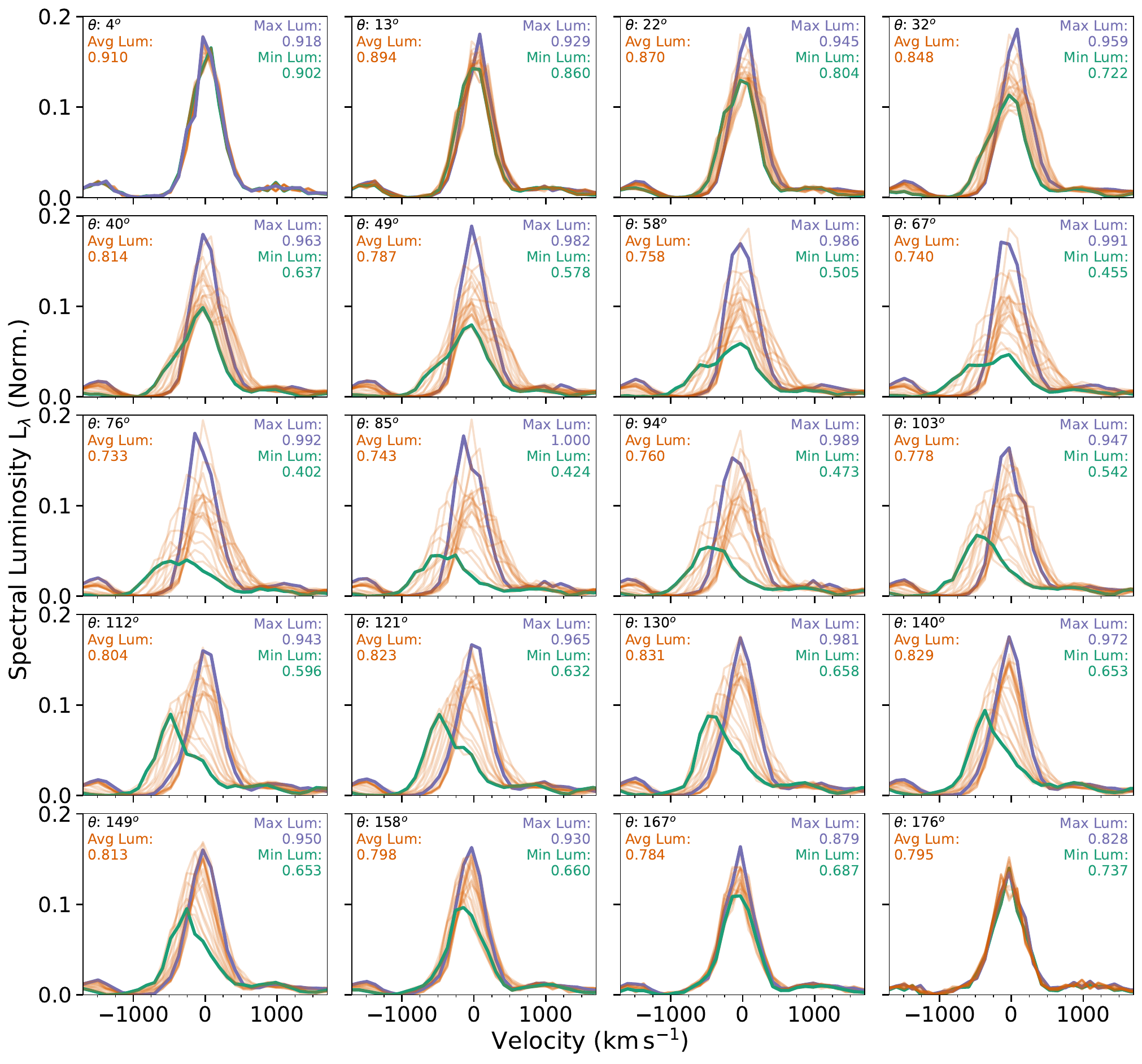}
    \caption{The same as Figure~\ref{fig:Halpha-400_fullVA}, but for Mg~I] $\lambda\,4571$, normalized relative to the most luminous Mg~I] profile (which is a different observer than in Figure~\ref{fig:Halpha-400_fullVA}).}
    \label{fig:MgI-400_fullVA}
\end{figure*}
Figures~\ref{fig:lineprofdet_sn1997D} and \ref{fig:lineprofdet_sn2016bkv} show that the models display significant variation in line profiles across the different viewing angles. In this section we take a more comprehensive look at this variation\txtred{, investigating the model at 400 days for various elements.}  %to investigate the variation of the line profiles within the model and to what degree this varies for the different elements. % We don't look at multiple lines from the same element.

In Figure~\ref{fig:Halpha-400_fullVA}, a full series of H$\alpha$ line profiles is shown for all $20\times20$ viewing angles, grouped per panel by the angle to the polar direction ($\theta$, denoted in the top left corner of each panel). Each profile is normalized to the luminosity of the brightest of these 400 profiles (which is the purple profile at \txtred{$\theta=121^\circ$}). For each panel, the brightest profile within that set is marked in purple, and the dimmest in green, with the remaining ones in orange. The viewing angles closest to the poles (min/max $\theta$ for north/south pole) have line profiles that are more similar to each other, as they are packed closer together than for the equatorial angles\footnote{This also means that these angles cover a smaller part of the sky, and this apparent size difference $\Delta\Omega_k$ is accounted for, see Section~\ref{sec:methods}.}.

\begin{figure*}
    \centering
    \includegraphics[width=\linewidth]{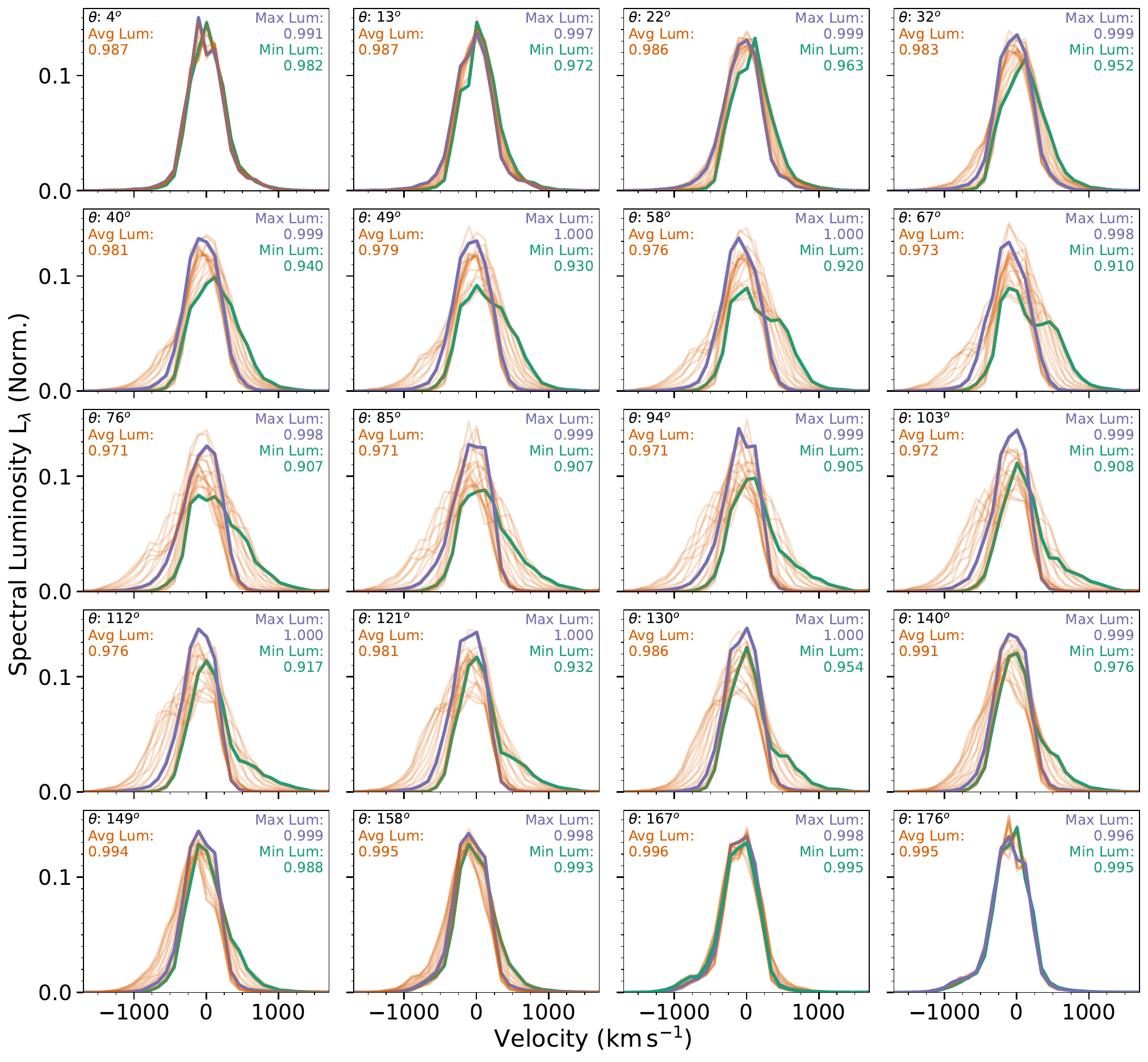}
    \caption{The [O~I] line profiles in s9.0 at 400 days, similar as in Figure~\ref{fig:Halpha-400_fullVA}. The profiles are each normalized to the most luminous [O~I] profile (the purple profile at $\theta=58^\circ$), accounting for both [O~I] features. For clarity, we only show the $6300\,\angstrom$ feature in this plot.}
    \label{fig:OI-400_fullVA}
\end{figure*}
One clear trend is that for each panel in Figure~\ref{fig:Halpha-400_fullVA}, the most luminous profile for a given $\theta$ is always on the blueshifted side (negative velocity, relative to H$\alpha$ rest frame), while the least luminous ones are (strongly) redshifted profiles. Overall, the least luminous H$\alpha$ profile is \txtred{$\sim84\,\%$} of the overall brightest one, which is purely due to radiative transfer effects. Additionally, the less luminous profiles for a given $\theta$ tend to be wider, which geometrically can be explained as having the stronger emitting regions more ``spread out'' across the line-of-sight for that viewing angle, rather than in the ``observing plane'', leading to wider profiles where the red sides are more strongly affected by radiation transport. Usually, less luminous profiles (roughly with luminosities below 90\% of the most luminous one) are also the more asymmetric profiles, as higher extinction across the line of sight leads to larger differences between the red and blue sides of the profile. Checking these profiles against Figure~\ref{fig:profile_props}, the broader profiles have more strongly shifted centroids.

In Figure~\ref{fig:MgI-400_fullVA}, the line profiles of Mg~I] $\lambda\,4571$ are shown, with the same structure as in Figure~\ref{fig:Halpha-400_fullVA}. The normalization for these profiles is also done to the most luminous of all 400 viewing angles, which is a different angle than for H$\alpha$ (with the brightest profile at \txtred{$\theta=85^\circ$} for Mg~I]). As Mg~I] is a much bluer line, the difference between the most luminous and least luminous profiles is also much bigger, because the line opacity increases significantly towards shorter wavelength \citep[e.g.][]{jerkstrand2015late}. Here the least luminous profiles are \txtred{not even} half as bright as the most luminous ones.

As with the H$\alpha$ profiles, usually the brightest profiles are the narrower ones, but, unlike H$\alpha$, the bright Mg~I] profiles tend to be (somewhat) redshifted and the least luminous profiles are blueshifted, or not shifted. In Figure~\ref{fig:profile_props}, the ``pattern'' between H$\alpha$ and Mg~I] \txtred{is} not very different although Mg~I] is more ``irregular''. From Figure~\ref{fig:profile_props}, Mg~I] does show a preference for blueshifted profiles on the wider end, which connects to these less luminous profiles here.

In Figure~\ref{fig:OI-400_fullVA}, we show the line profile variations for all viewing angles for the [O~I] $\lambda\lambda\,6300,\,6364$ line, for the pure O-emission, with the same structure as in Figure~\ref{fig:Halpha-400_fullVA}. The luminosities L$_\lambda$ are normalized using the full doublet, although only the profiles of the $6300\,\angstrom$ feature are shown to better highlight the different profile shapes. The brightest profile for [O~I] is the purple profile at $\theta=58^\circ$. The difference between the brightest and dimmest profile is \txtred{$\sim10\,\%$}, \txtred{somewhat less than for} H$\alpha$, \txtred{and} the average luminosities (per $\theta$) are \txtred{also} slightly higher for [O~I] than they were for H$\alpha$.

\begin{figure*}
    \centering
    \includegraphics[width=\linewidth]{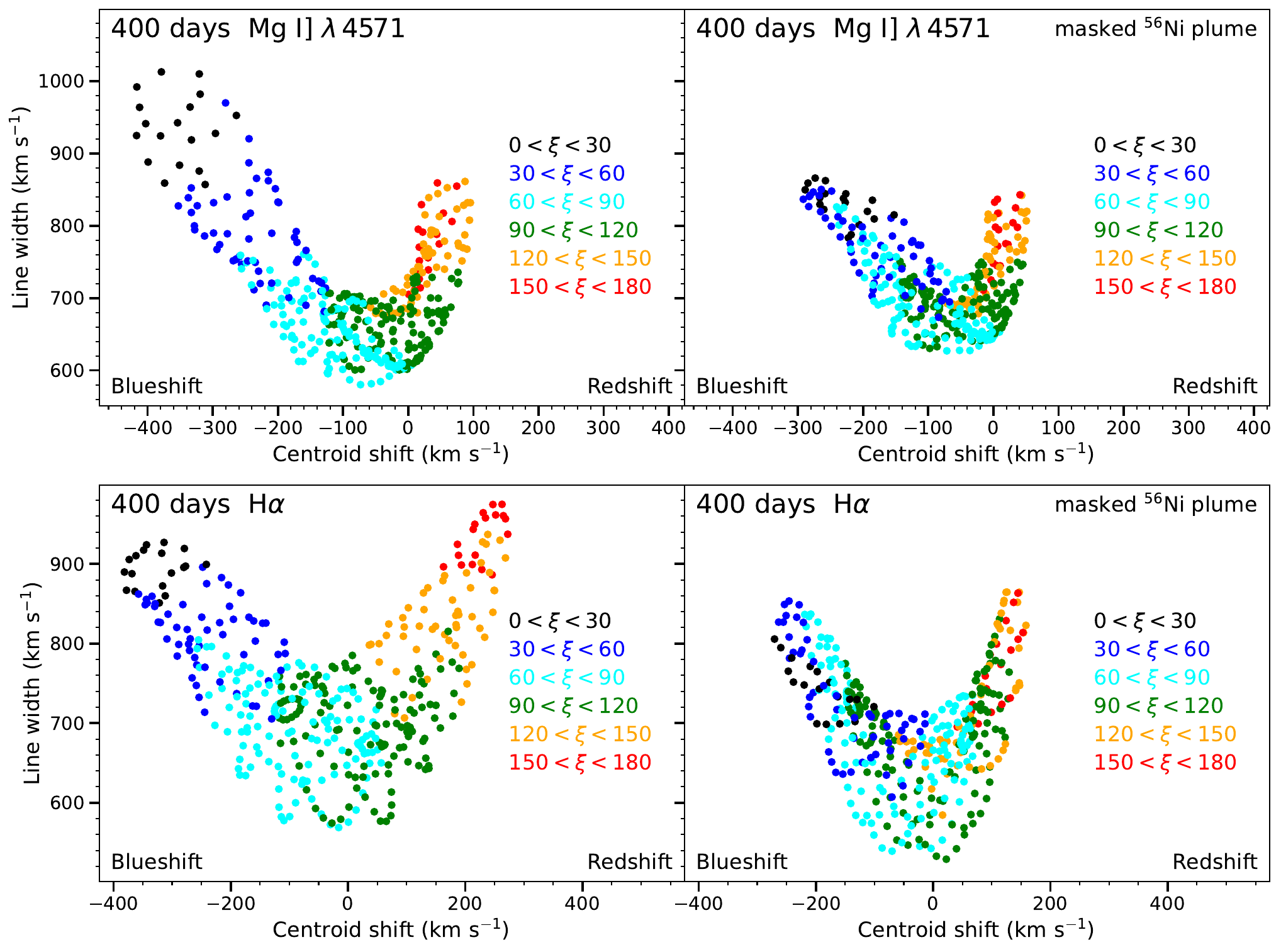}
    \caption{\txtred{The centroid shifts and line widths for Mg~I] (top) and H$\alpha$ (bottom), colour coded by the angle $\xi$, which is the angle between direction vector to the observer and the direction of the main $^{56}$Ni plume (red cross in Figure~\ref{fig:Niplume-aitoff}). $\xi = 0$ means that the centre of the Ni plume points exactly to the observer. On the left, the line profile properties for the regular setup are shown, while on the right they are computed from a setup where the $\gamma$-ray energy deposition from the primary $^{56}$Ni plume is masked (see text). On both rows, both panels have the same axis range to better highlight the impact of this masking.}}
    \label{fig:NiPlume_shiftwidth}
\end{figure*}
Between Figures~\ref{fig:Halpha-400_fullVA}, \ref{fig:MgI-400_fullVA} and \ref{fig:OI-400_fullVA}, it becomes clear that the line profiles across different viewing angles are unique to each element (and potentially even specific ion species). This occurs even when elements have, broadly speaking, similar line profile properties like H$\alpha$ and [O~I] ($v_\text{shift}$ and $v_\text{width}$, see Figure~\ref{fig:profile_props}). Despite the modest explosion energy in s9.0, large variations appear due to radiative transfer effects, with the different lines each impacted in their own unique way. Mg~I] has a blue wavelength ($4571\,\angstrom$), and thus is more impacted by radiative transfer than H$\alpha$ or [O~I]. However, even for very red lines like [C~I] $\lambda\,8727$, these effects can be of similar size as for H$\alpha$.

\txtred{To showcase that the radiative transfer effects are stronger at early times, and thus that the variations with viewing angle are larger, in Appendix~\ref{app:OI250day} (Figure~\ref{fig:OI-250_fullVA}) the [O~I] line profiles are shown at 250 days. While in each $\theta$ set, the most luminous profiles (the purple profiles) are already very close in luminosity to the overall most luminous profile (at $\theta=121^\circ$), the least luminous lines at 250 days are are missing about 1/3rd of the total luminosity. As at 400 days for H$\alpha$ and [O~I], the least luminous profiles are more redshifted and less symmetric. There are also several profiles (at e.g. $\theta=76^\circ$) which have noticeable emission bumps around \kms{-1000}, while at 400 days none of the three elements display such an effect, indicating that over time the emission becomes more confined to the central regions.}

\subsection{\txtred{$^{56}$Ni plume alignment}} \label{ssec:NiPlume}
\subsubsection{\txtred{Line profile properties}}
\txtred{One of the main characteristics of the input model is the presence of a prominent $^{56}$Ni plume ejected at a velocity up to $\sim$\,1500\,km\,s$^{-1}$ (red cross in Figure~\ref{fig:Niplume-aitoff}). As the direction vector of this plume is noticeably offset from the total neutron star kick (gray cross in Figure~\ref{fig:Niplume-aitoff}), we also compare here how the line profile properties vary by observer for the angle $\xi$, which is the angle between the observer and this prominent Ni plume. In the left panels of Figure~\ref{fig:NiPlume_shiftwidth}, the properties for Mg~I] (top) and H$\alpha$ (bottom) are shown, colour coded by $\xi$.} %% \txtred{In Appendix~\ref{app:NiPlumeApp} (Figure~\ref{fig:NiPlume_props}), we show all six features (Mg~I], [O~I], H$\alpha$, [Fe~II], [Ca~II] and [C~I]).}

\txtred{The calculations of $v_\text{shift}$ and $v_\text{width}$ have not been changed, so only the colours vary compared to Figure~\ref{fig:profile_props}. What can clearly be noticed here in Figure~\ref{fig:NiPlume_shiftwidth} is that the observers best aligned with the Ni plume, with the plume pointing to them (black dots), always see the most blueshifted and widest profiles for both lines. Additionally, the most anti-aligned observers (red dots), always see the most redshifted profiles. Specifically for Mg~I], the anti-aligned observers also tend to see somewhat more narrow profiles than their aligned counterparts, which does not occur for H$\alpha$.}

\txtred{In order to better investigate the impact of the primary Ni plume in the model, we also ran the model while masking this $^{56}$Ni plume. This masking was done by preventing the cells that contain the $^{56}$Ni plume from generating their $\gamma$-rays, while $\gamma$-rays that originate elsewhere can still enter the region and deposit their energy. This plume holds about $25\,\%$ of all $^{56}$Ni, and as a result of this masking the total $\gamma$-ray deposition drops to $1.57\times10^{39}\,$erg (about $80\,\%$ of the total in the normal setup). For the spectral computations, this region is still fully accounted for, but now holds far less energy than before. In the right panels of Figure~\ref{fig:NiPlume_shiftwidth}, we show the line profile properties of Mg~I] and H$\alpha$ for this masked setup, colour coded by $\xi$.} 

\txtred{Generally, what can be seen is that when the $^{56}$Ni plume is masked, the centroids of the lines are less shifted, and the maximum values of the line widths also decrease somewhat, with Mg~I] decreasing to line widths of at most \kms{875} (down from \kms{1000}) and H$\alpha$ to \kms{875} (from \kms{975}). For Mg~I], the line centroids are now confined between \kms{-300\text{ and }50} and for H$\alpha$ between \kms{-300\text{ and }150}. It can also be seen that the most extreme profiles no longer neatly align with angle $\xi$, i.e. the direction of the primary $^{56}$Ni plume. This can directly be attributed to the loss of the $\gamma$-rays from the primary $^{56}$Ni plume, although there still is a reasonable separation by colours in Figure~\ref{fig:NiPlume_shiftwidth} for the masked setup. The primary $^{56}$Ni plume leads to a direct impact on the nebular phase through the more extreme shifting of the line centroid and further broadening of the total line.}

\subsubsection{\txtred{Line profile details}}
\txtred{A more direct test of the impact of the $^{56}$Ni plume on the line profiles can be made by directly comparing the line profiles of the normal setup with the line profiles in the masked setup. For this comparison, we focus on H$\alpha$, [Ca~II] $\lambda\lambda\,7291,\,7323$ and Mg~I] $\lambda\,4571$, which are the two strongest uncontaminated lines (H, Ca) and the bluest line (Mg) which should have the largest radiative transfer impact.}

\begin{figure*}
    \centering
    \includegraphics[width=.495\linewidth]{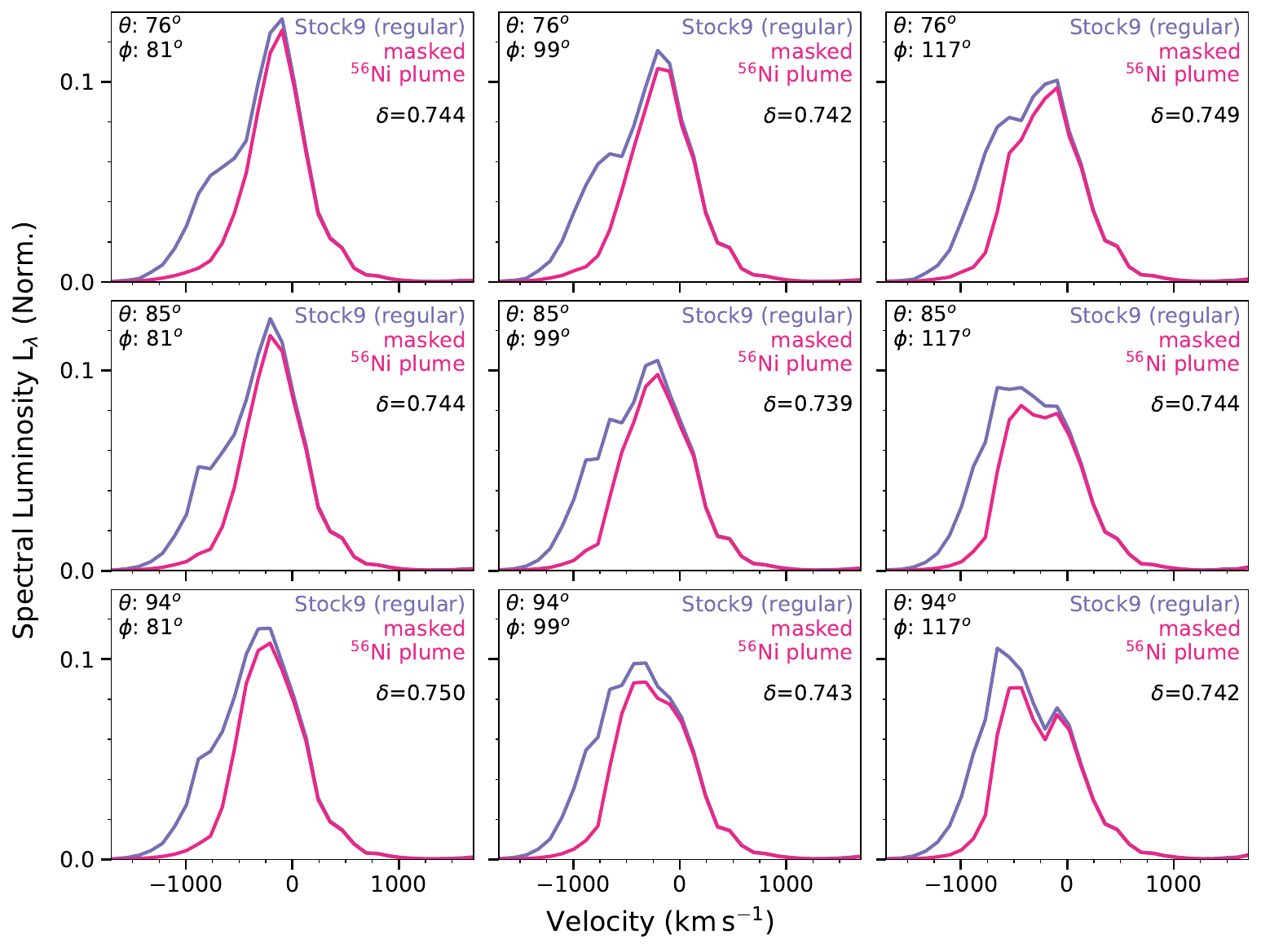}
    \vspace*{.5cm}  % To add a small bit of whitespace between the element-plots
    \includegraphics[width=.495\linewidth]{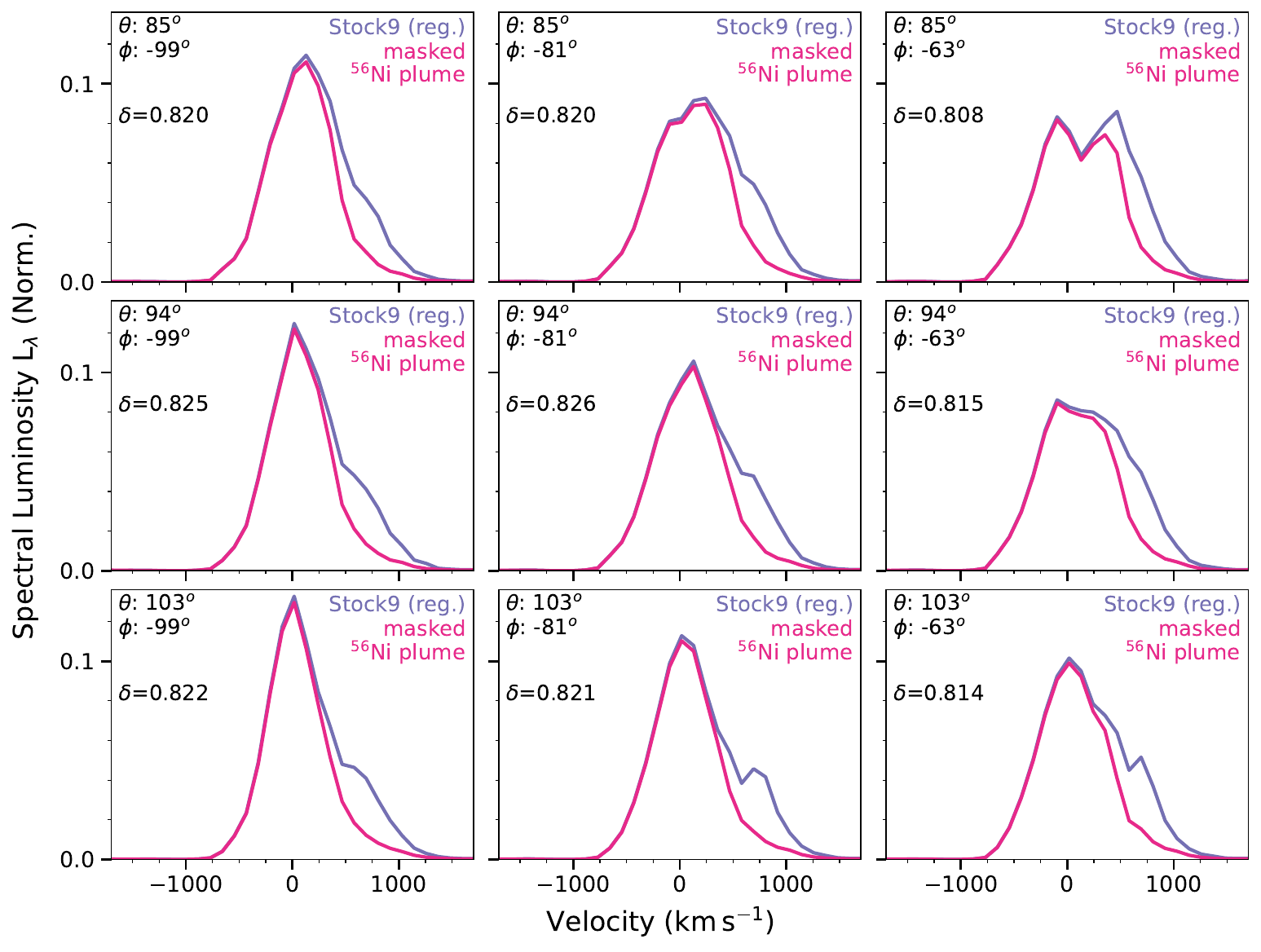}
    \includegraphics[width=.495\linewidth]{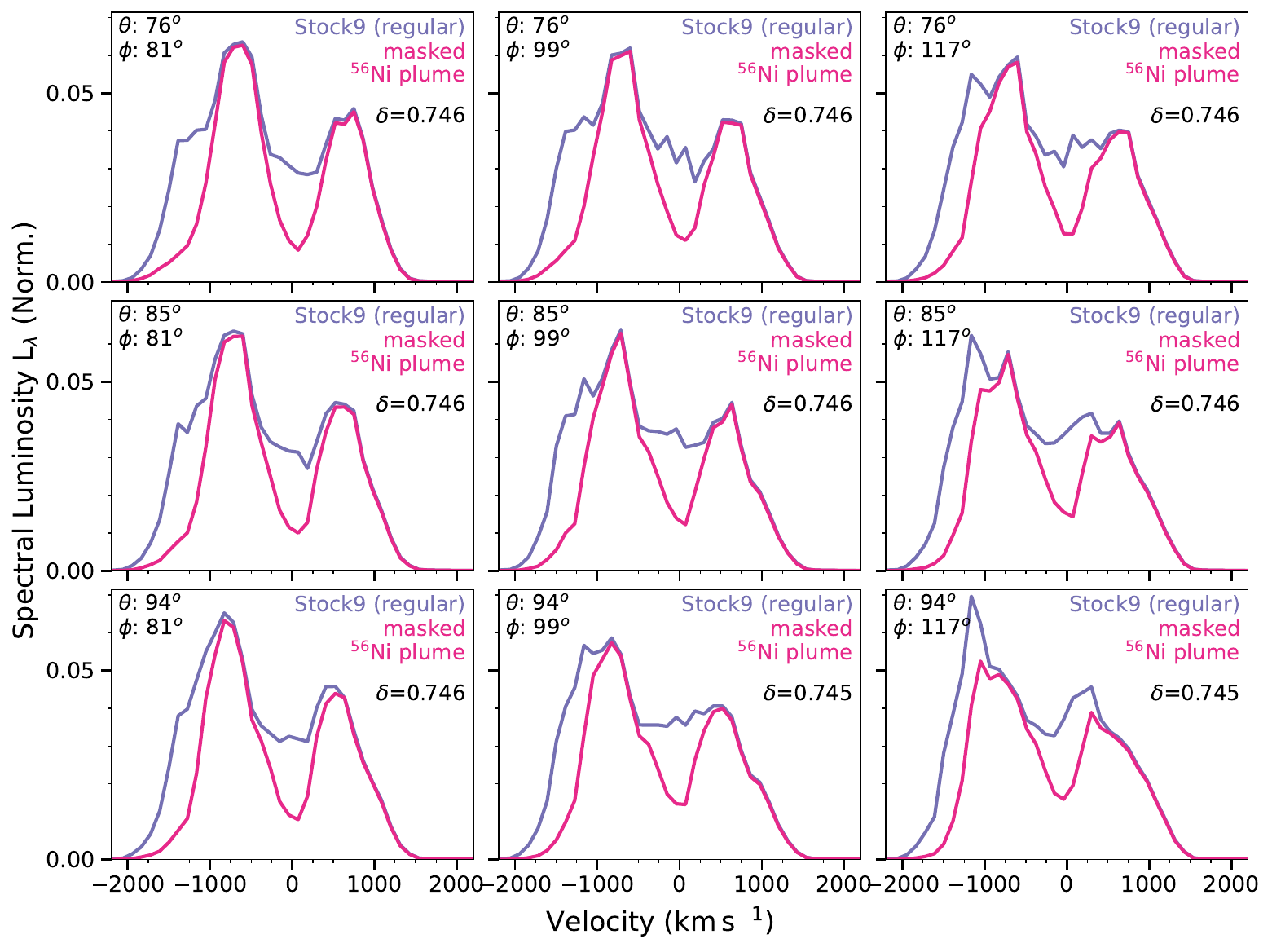}
    \vspace*{.5cm}  % To add a small bit of whitespace between the element-plots
    \includegraphics[width=.495\linewidth]{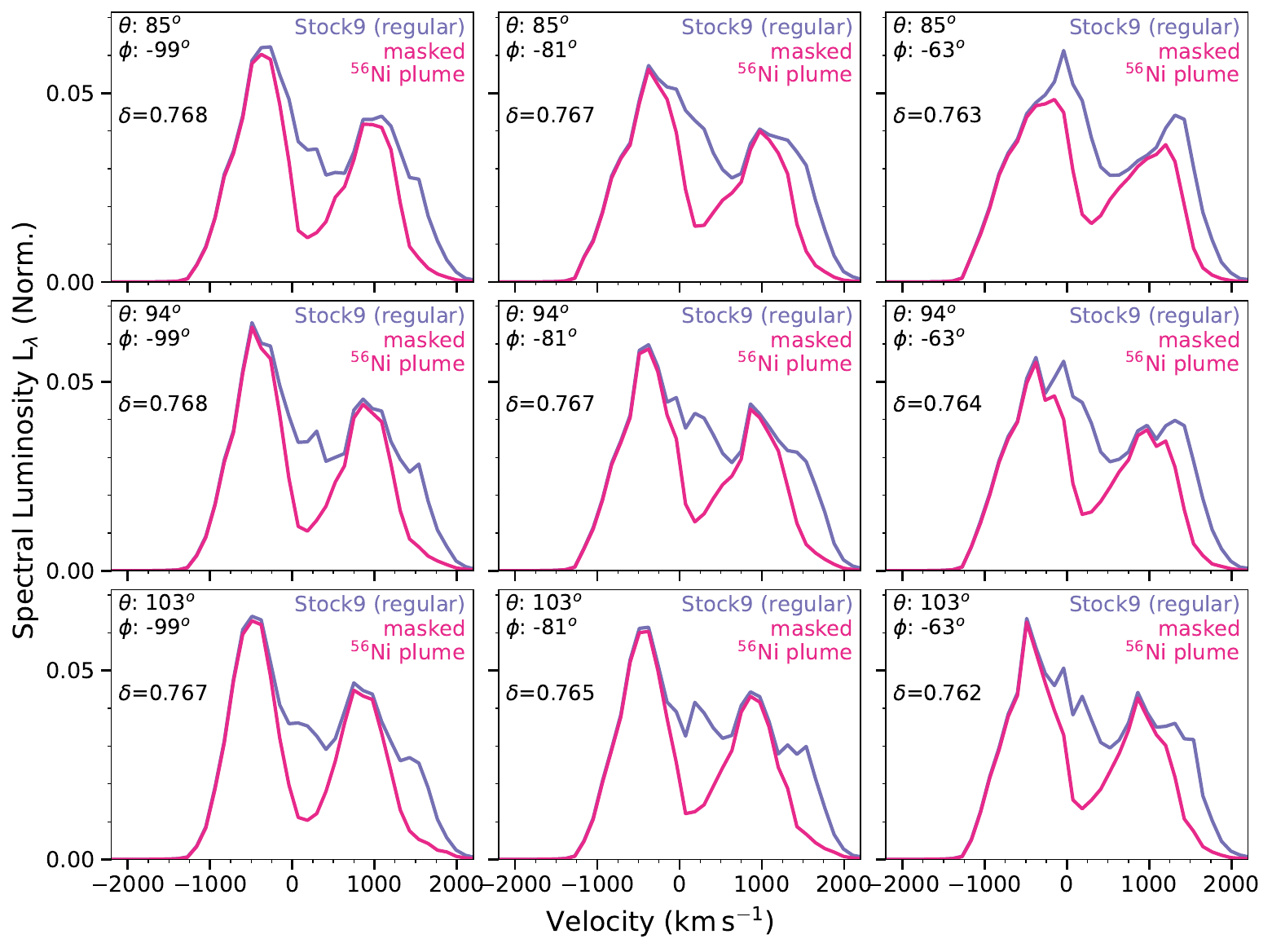}
    \includegraphics[width=.495\linewidth]{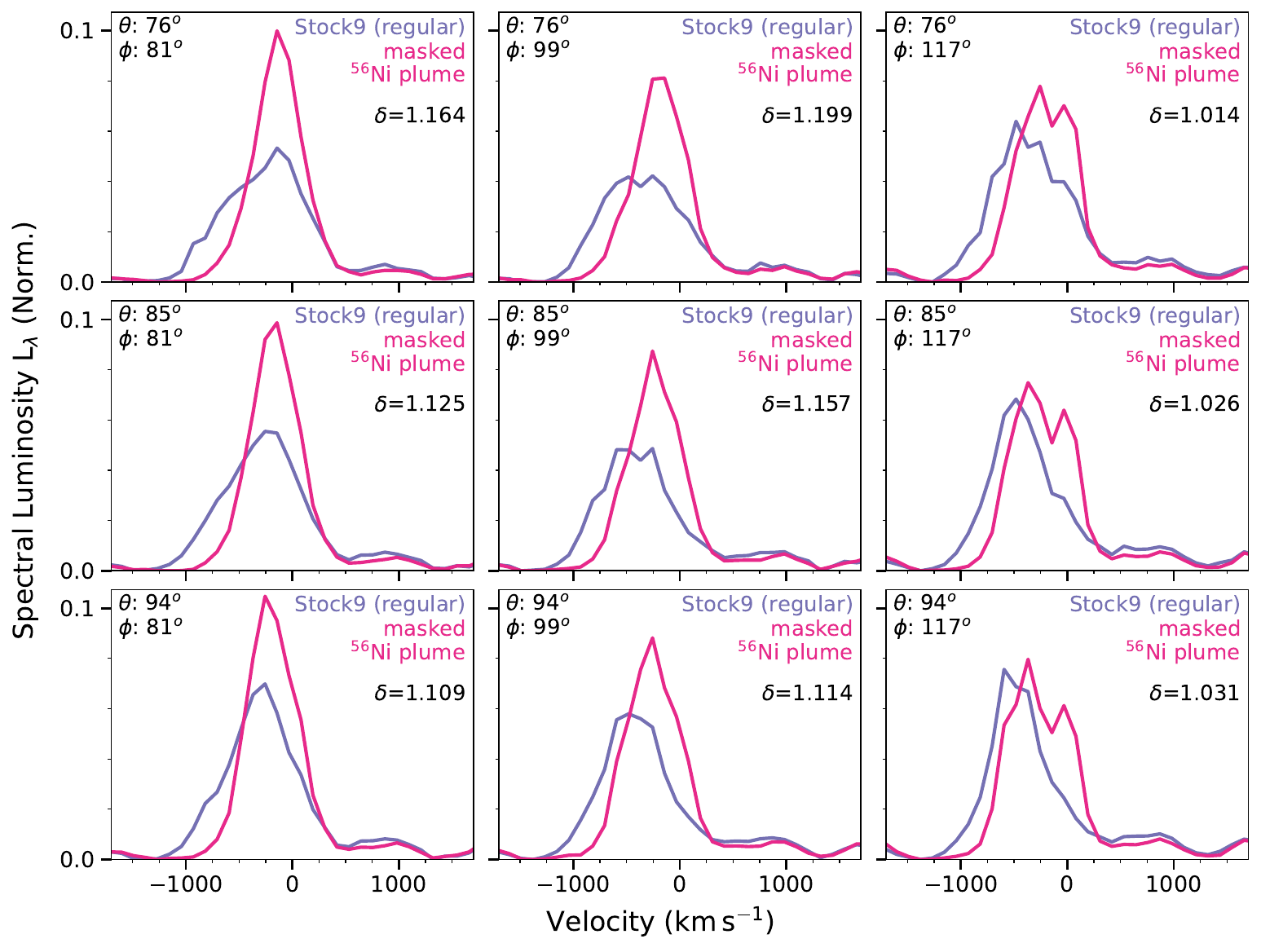}
    \includegraphics[width=.495\linewidth]{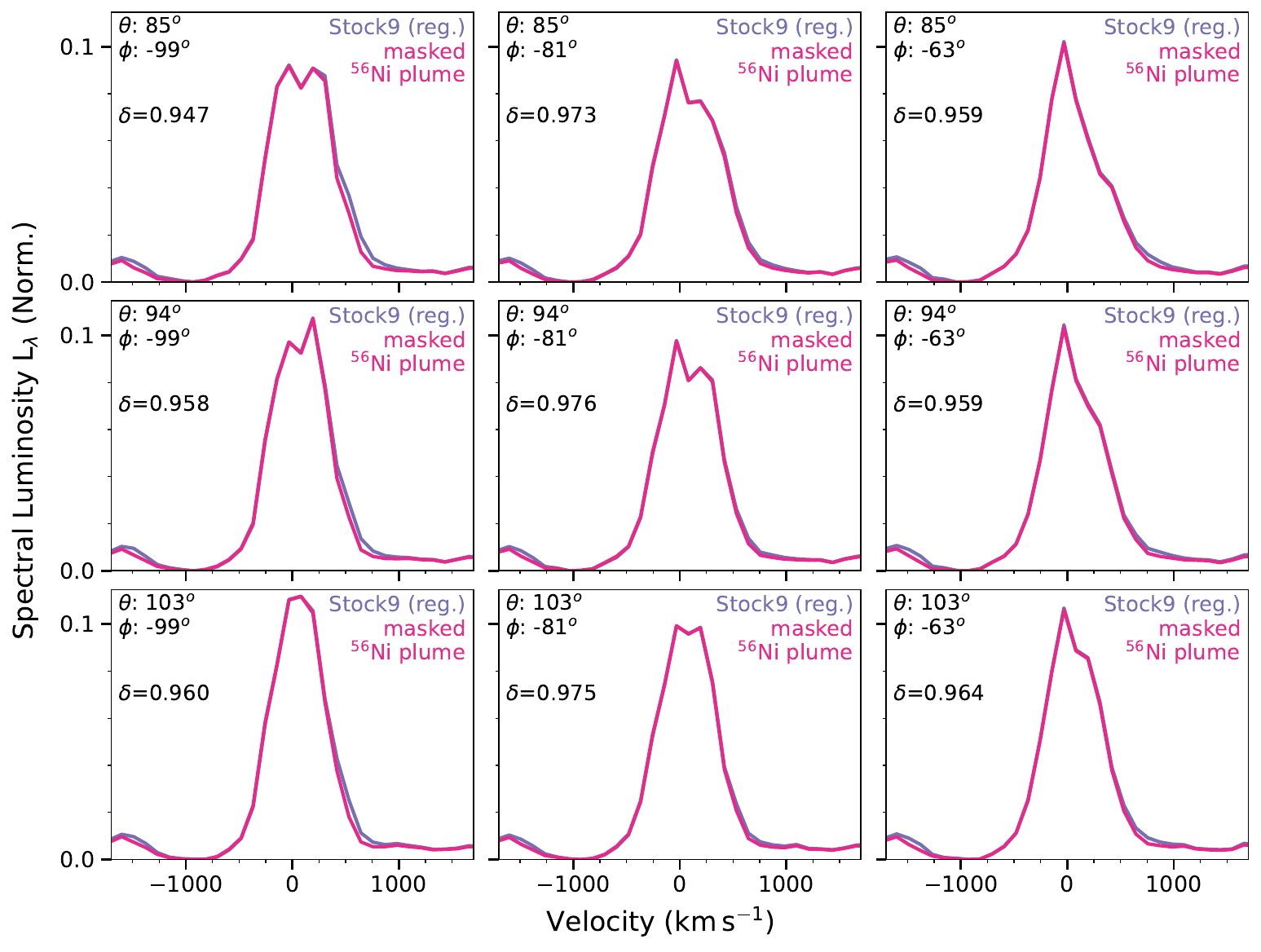}
    \caption{\txtred{A $3\times3$ panel setup for the line profiles of H$\alpha$ (top), [Ca~II] $\lambda\lambda\,7291,\,7323$ (middle, with the line centre on $7304\,\angstrom$), and Mg~I] $\lambda\,4571$ (bottom) containing the nine observers best aligning with the $^{56}$Ni plume (left column) and the nine observers most anti-aligned with this plume (right column) -- the detailed $\theta$ and $\phi$ coordinates of the viewers are denoted in each separate panel. In each panel, the line profiles of the normal setup (in purple) are shown alongside the line profiles of the masked $^{56}$Ni plume setup (in pink), both normalized to the most luminous profile among all viewers across both setups. The ratio $\delta$ indicates the ratio of the integrated line luminosity of the masked setup divided by the normal setup.}}
    \label{fig:NiPlumeProfiles_full}
\end{figure*}
\txtred{In Figure~\ref{fig:NiPlumeProfiles_full}, we showcase the normal (purple) and masked $^{56}$Ni plume (pink) models for the three aforementioned lines (H$\alpha$, [Ca~II], Mg~I], from top to bottom), for the nine viewers best aligned with the primary $^{56}$Ni plume (left column) and the nine viewers most anti-aligned to the $^{56}$Ni plume (right column), and denote the ratio of the integrated line luminosities between the models as $\delta = L_\mathrm{masked} / L_\mathrm{regular}$, such that $\delta<1$ indicates observers for whom the regular setup is more luminous.}

\txtred{For H$\alpha$, it can be seen that when going to the masked setups, the bump on the blue (red) side disappears for the best (worst) aligning observers, which corresponds to the emission from the $^{56}$Ni plume region which disappears in the masked setup, as the total deposition in that region is much lower. The peaks of the line profiles are barely affected, but this does lead to more narrow profiles in the masked setup, which explains the decrease in line widths and centroid shifts already noted in Figure~\ref{fig:NiPlume_shiftwidth}, between the regular and masked setups. The ratio $\delta$ for the best-aligning viewers is $\sim0.75$, which matches the drop in total energy deposition for this model. For the worst-aligning viewers, the ratio $\delta$ is somewhat different at $\sim0.82$, indicating that in the normal setup some fraction of the total emitted H$\alpha$ does not reach these observers due to radiative transfer effects.}

\txtred{For [Ca~II], the pattern in the best-aligning observer directions is much the same as for H$\alpha$, where the blue components of the line profiles are largely missing in the masked setup (almost leading to narrow enough profiles that the line doublet is no longer blended), and the ratio $\delta$ is also $\sim0.75$ for [Ca~II]. For the anti-aligned viewers, the missing emission resides on the red side of the profiles, but for [Ca~II] the ratio $\delta$ is almost the same for both aligned and anti-aligned observers. This indicates that the radiative transfer effects connected to the masking of the primary $^{56}$Ni plume have only a very minor impact on the line luminosity of [Ca~II].}  % specifically the TRANSFER effects are small, of course the overall luminosity drop is not small!

\txtred{For Mg~I], a completely different trend appears. Starting with the anti-aligned observers, it can be seen that the ratio $\delta$ between the models is close to 1, indicating that the loss of the energy deposition in the primary $^{56}$Ni plume is barely noticed by the viewers located on the opposite side of the nebula, with the line profiles barely deviating. For the aligned observers, a strong effect is noticeable and with something quite remarkable: $\delta$ is greater than 1, indicating that the Mg~I] profiles in the masked setup are more luminous than they were in the normal setup, even though the normal setup has a higher total energy deposition. The blue shoulder in the line profiles still disappears for the masked setup (note the lack of emission in the pink profiles at \kms{v\sim1000}) but the central component of the lines becomes much stronger instead, creating profiles which are more luminous by up to about $15\,\%$.}

\txtred{The behaviour of the Mg~I] line profiles for the best-aligning viewers is quite surprising, especially given that H$\alpha$ and [Ca~II] behave differently (although also for these, the bluest component in their profiles disappears). The appearance of the extra central emission for these viewers -- the main cause as to why $\delta$ exceeds 1 for these viewers -- could either originate from enhanced central Mg~I] emission in the masked model, or from the disappearance of nearby blocking lines (Fe, Co, Cr and Ti all have lines in the region $4571-4590\,\angstrom$ which could absorb Mg~I]). However, the anti-aligned viewers give no indication of increased central Mg~I] emission in the masked setup, making the disappearance of blocking lines a more likely explanation.}

\begin{figure}
    \centering
    \includegraphics[width=\linewidth]{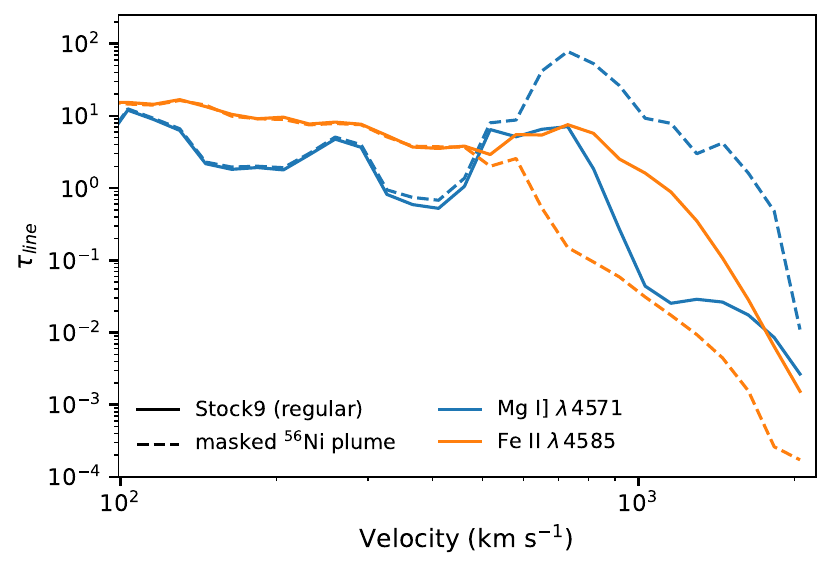}
    \caption{\txtred{A comparison between the line opacities of Mg~I] $\lambda\,4571$ (blue) and Fe~II $\lambda\,4585$ (orange) for the normal setup (solid lines) and the model where the primary $^{56}$Ni plume is masked (dashed lines), focused on the cells inside the cone that contain the $^{56}$Ni plume.}}
    \label{fig:MgFe_taucomp}
\end{figure}

\txtred{To investigate this, in Figure~\ref{fig:MgFe_taucomp} we showcase the optical depth $\tau_\text{line}$ for Mg~I] (blue) and one of these potential blocking lines, Fe~II $\lambda\,4585$ (orange), for the $\theta$ and $\phi$ cone of the grid which contain the $^{56}$Ni plume and which thus contain the masked region specifically, for both models. What can be seen is that up to \kms{v\approx500}, the optical depths of the lines are effectively identical between the two models, but at higher velocities -- where the masking took place, and thus the $\gamma$-ray deposition is significantly altered -- the two lines react in a completely different manner to the change in $\gamma$-ray deposition. The Mg~I] line optical depth increases by more than an order of magnitude, while the Fe~II one drops by almost two orders of magnitude. The stark increase in $\tau$ for Mg~I] leads to the reduction of Mg~I emission at these higher velocities, explaining the drop in the blueshifted component of the line profile as that emission is now largely prevented. At the same time, the huge drop in $\tau_\text{line}$ for this Fe~II line indicates that this line will no longer have a noticeable impact as line blocker, enabling more central Mg~I] emission to leak out towards the observers on that side.}

\section{Discussion} \label{sec:discussion}
One of the big questions we can address with 3D modelling is to what degree line profiles vary with viewing angle, both regarding line strength and in line profile. Getting a good grasp on the intrinsic variability due to the asymmetric nature of the explosion is important, as line strengths are used to infer ejecta masses of key elements (e.g. oxygen through the [O~I] $\lambda\lambda\,6300,\,6364$ doublet, \citealt{jerkstrand2014nebular,jerkstrand2015late}). In the s9.0 model, we find that the [O~I] luminosity might vary by \txtred{$\sim10\%$} between the most luminous and least luminous profiles (see Figure~\ref{fig:OI-400_fullVA}), depending on the viewing angle\txtred{, and that this effect is much bigger at earlier times (up to $\sim33\,\%$, see Figure~\ref{fig:OI-250_fullVA})}. It is unclear how this translates to other models and if such viewing angle effects are of similar size, but if they are then it complicates ejecta composition inferences using 1D modelling. Furthermore, different elements have different variations with viewing angles, so even using ratios between different lines will not be a consistent alternative.

The model used in this work can be compared to observed LLIIP SNe due to its low expansion velocity and ejecta mass. When comparing the line profiles from the model to two such SNe (SN~1997D and SN~2016bkv) we find that the strong lines in the observed spectra can be well-matched for some viewing angles. It should however be noted that, for the most part, the line ratios between these lines are still not fully comparable between the model and the observations, as our model spectra have particularly strong H$\alpha$. However, as shown in Figure~\ref{fig:PIB-SN1997D}, outside of H$\alpha$ there are reasonable matches to most line luminosities of SN~1997D, at both 325 and 459 days. For SN~2016bkv this comparison is harder to make, as there is a strong blue contaminating component with a slope into the red that is very difficult to reliably remove. This contamination prevents a comparison between normalized spectra. Despite this contamination however, line profiles of individual lines can be compared with our model.

Overall, the model here can give good fits to many different line profiles for SN~1997D and SN~2016bkv for the same viewing angle, although there is not one angle where all five investigated profiles (Figures~\ref{fig:lineprofdet_sn1997D} and \ref{fig:lineprofdet_sn2016bkv}) give good matches. However, individually each profile has good matches for some viewing angles and the best angles overall tend to find good fits for three or four lines at the same time. This is strongly encouraging for the potential diagnostic use of 3D explosion models processed by nebular phase spectral codes. This is particularly so because this is not a model tailored to match any specific SN, making the level of agreement shown here to be satisfactory alongside the variations of the line profile properties (line centroid shift and line width) for the different viewing angles with regards to the model.

% the model presented is for an iron-CCSN and it is stated that the models does match the observed profiles of SN2016bkv well but this is used to support the conclusions that SN2016bkv is actually an EC supernova. I think the point being made is that the 3D results here are not sufficiently different from previous 1D studies that the conclusions of those 1D studies should be revised - but some rephrasing might help to make this clearer, if so.
SN~2016bkv was tentatively suggested to be of ECSN origin. The line profile comparisons in Figure~\ref{fig:lineprofdet_sn2016bkv} do find that our model \txtred{-- an Fe-CCSN --} can match the observed profiles well, at varying viewing angles for the different elements. Although we do not directly compare the full model spectrum against SN~2016bkv, we have similar luminosities for the key lines as in the analysis by \citet{hosseinzadeh2018short}.
\txtred{As our results here are not sufficiently different from a previous 1D study (\J18), there is no reason to cast doubts on the viability of 1D modelling for low-energy SNe. Therefore the conclusions drawn by \citet{hosseinzadeh2018short} on grounds of their 1D modelling cannot be refuted, namely that SN~2016bkv could be an ECSN based on nebular phase models, although it has an unusually large $^{56}$Ni mass. However, according to our results an Fe-CCSN appears to be a possible scenario as well.}
Another ECSN candidate is SN~2018zd, which we do not compare to here -- the main reason is that the features in SN~2018zd are either missing from the observations (Mg~I]) or much broader (H$\alpha$ and [O~I] in particular) than in our models.

From the line profile comparisons between our model and SN~1997D and SN~2016bkv, is should be noted that most viewing angles in the model struggle to get a good match to Mg~I] $\lambda\,4571$, in particular due to the extended red side of the observed profiles. Conversely, there are a lot of [Ca~II] $\lambda\lambda7291,\,7323$ profiles which find exceptionally good matches to both observed SNe. For the other three investigated lines, it is harder to get exceptionally good matches although all of them find reasonably good comparisons at some angle. This is likely due to the fact that bluer wavelengths are more strongly affected by radiative transfer, while the [Ca~II] profile only suffers mildly. This is a tentative suggestion that the degree to which the Mg~I] profile can match e.g. the [O~I] profile does not intrinsically depend on the progenitor star, but rather the geometry of the explosion and our subsequent view thereof.

One area of major difference between the models and the observations is the appearance of several strong Fe~I lines in the model (e.g. Fe~I $5957\,\angstrom$ and near $6364\,\angstrom$) which are not present, or only weakly present, in the observations. This discrepancy may indicate too high densities in the model used here, which leads to strong recombination of Fe~II into Fe~I and therefore a quite neutral iron ionization balance, which then impacts the overall spectral synthesis in several ways. \txtred{For the comparison to \J18 shown in Figure~\ref{fig:SUMO-ExTraSS_spectral}, the missing Fe~II emission at $4000$--$4500\,\angstrom$ and $5000$--$5500\,\angstrom$ instead seems to appear as Fe~I emission around $6200$--$6700\,\angstrom$ and $7900$--$8400\,\angstrom$.}

The exact cause for the particularly strong Fe~I emission is unknown. Aside from the higher densities leading to more recombination, it can be seen in Figure~\ref{fig:composition_stock} that pretty much everywhere in the ejecta, the mass fraction of Fe is approximately $10^{-3}$ or higher. This means that also at higher velocities, there is Fe which can participate in the emission (Fe~I is an efficient cooler). This can in turn drive down temperature and subsequently further alter the physical conditions. It is unclear if the enhanced mixing of the $^{56}$Ni into the envelope, where primordial Fe does some of the cooling, also aids in powering the Fe~I emission, or if this emission is enhanced due to the small nuclear network used in the explosion model. The small network leads to an overabundance of $^{56}$Ni \citep[see e.g.][Sec 3.4, who discuss this overproduction]{bollig2021self}. \citet{kozyreva2022low} used a model with a $^{56}$Ni mass of $0.003\,M_\odot$, which is half of what we use here but which still gives good matches to LLIIP SN light curves. An overproduction of $^{56}$Ni leads to more $^{56}$Fe at nebular times, which could be a reason for our strong Fe~I emission.

% \txtred{NOTE this paragraph needs to be re-written -- the difference in PI levels between SUMO and ExTraSS is much less severe now. Also no longer need the boosted PI rates.}
% We investigated if a boosted photoionization rate (by increasing $\sigma_\text{PI}$ by a factor 1000 for Fe and Ca) could correct the strong Fe~I and Ca~I emission. We found that this does largely alleviate the issue of the Ca~I $\lambda\,4226$ emission and it somewhat reduces Fe~I emission, in particular at $5957\,\angstrom$, but Fe~I remains a strong emitter throughout the ejecta. An alternative strategy could be to use more states in our photoionization calculations, as we only included up to level 13, while in \texttt{SUMO} (and thus in \J18) up to level 50 is considered. A large fraction of the unexpected Fe~I emission originates from the 5D z7P states, which are levels 44, 47 and 50, and thus are considered for photoionization by \J18 but not here.
\txtred{Alongside the strong Fe~I emission, our model here also finds very strong H$\alpha$ emission compared to both \J18 and the observations. The model here has slightly higher Fe in the envelope and we do not adjust for the solar composition here either. This means that we cannot reproduce lines of e.g. stable Ni ([Ni~II] $\lambda\,7378$) or Na (Na~I D), which could have carried some of the excess cooling that H$\alpha$ and Fe~I might be doing instead here. In \citet{vanbaal2025ExTraSS} a direct comparison between the \J18 model and a 3D version thereof is made, where the solar composition is accounted for, and they find that this helps reduce the H$\alpha$ and Fe~I emission significantly, indicating that the adjustment for solar composition and/or density effects are largely responsible for this excess emission here.}
% \txtred{Discuss instead the lack of solar-composition correction, to add N/Na/Ni etc?}

Although the model here is downsampled even more than in the grids of \citet{vanbaal2023modelling,vanbaal2024diagnostics}, enough resolution ($\Delta v/v$) is retained for the spectral synthesis. The last radius in our model covers the ejecta with \kms{1830\leqslant v_\text{ej} \leqslant2050}, yet contains only $0.01\,M_\odot$ ejecta. The angular downsizing was checked, but increasing the angular resolution back to $30\times60$ did not have any significant impact in the spectral synthesis and line profile generation, but does come at a significantly higher computational cost.

It should be noted that the explosion energy in our model here is around a factor 2 lower than in the 1D model of \J18. The explosion energy range of observed LLIIP SNe was put at $0.1-0.28\times10^{51}\,$erg (median $0.17^{+0.07}_{-0.03}\times10^{51}\,$erg) by \citet{das2026low}, using models from \citet{moriya2023synthetic}. The model used here exploded with $0.054\times10^{51}\,$erg \citep{janka2024interplay}, which is lower compared to what is estimated for the observed sample. \citet{kozyreva2022low} found, however, good multi-band light curve matches with the same s9.0 star at an explosion energy of $0.068\times10^{51}\,$erg when comparing to SNe 2005cs and 2020cxd. A result of this low explosion energy is that the densities in our model may be too high (for a given epoch), which results in particularly strong H$\alpha$ and Fe~I emission. In \J18, the model line widths were $46\,\%$ larger than observed at 350 days for SN~1997D (see their Table 3), while here, when taking viewing angle \#165 (the best overall angle in the model at 400 days, see Figure~\ref{fig:lineprofdet_sn1997D}) gives line widths $20\,\%$ smaller than observed at 459 days. % I still think we shouldn't outright state "this means that the explosion energy lies somewhere inbetween, since it could also be due to lower ejecta mass compensating the line widths.

A higher explosion energy would mean higher expansion velocities and therefore lower densities for a given epoch, which would likely lead to weaker H$\alpha$ and Fe~I emission, although it would also impact the line widths for which we do find good matches against SN~1997D and SN~2016bkv. Additionally, a higher explosion energy would (likely) lead to a higher $^{56}$Ni mass, which would lead to even more synthesized Fe (after the $^{56}$Ni and $^{56}$Co decay) and thus it is not obvious if a higher explosion energy would resolve the H$\alpha$ and Fe~I emission strengths. Having a somewhat lower ejecta mass with the same explosion energy would also lead to higher expansion velocities, and the partial loss of the H envelope would additionally also lead to weaker H$\alpha$ emission. Mass loss and pulsations \citep[see e.g.][]{goldberg2020massive,bronner2025explosions,laplace2026pulsations} in the late stages of stars are poorly understood, and can impact the light curve significantly, complicating explosion energy estimates. % Mass loss, leading to a partial stripping of the H-rich envelope, could also occur through binary interactions.

\txtred{When comparing the alignment between the observers and the neutron star kick direction ($\Psi$, as used in Figure~\ref{fig:profile_props}), we find marginal separation by this angle $\Psi$, indicating that the main imprints of asymmetry in the model do not align with this kick vector. This can be understood by the fact that the neutron star kick in this model is small and strongly dominated by the neutrino-induced kick instead of the ejecta-connected hydrodynamic kick. As shown in Figure~\ref{fig:NiPlume_shiftwidth}, the correlation between $\xi$ (the angle between the observer and the fast-moving, primary $^{56}$Ni plume, see also Figure~\ref{fig:Niplume-aitoff}) and the line profile properties is much stronger. We find that angles most aligned to this plume (plume pointing to observer) show the most blueshifted and wide profiles, while anti-aligned observers instead see the reddest profiles, and in the case of Mg~I] also somewhat less wide profiles. This alignment directly indicates that the asymmetry, imprinted with the Ni plume, leads to observational differences for the line profiles.}

\txtred{To more directly test the impact of the primary $^{56}$Ni plume, we also ran a modified setup where the $
\gamma$-ray energy deposition from this primary plume specifically was masked (but $\gamma$-rays from elsewhere can still deposit their energy if they pierce into the masked region). Figure~\ref{fig:NiPlume_shiftwidth} shows the impact on the line profile properties for Mg~I] and H$\alpha$ between the regular setup and the masked setup: the correlation with $\xi$ decreases, and variation with viewing angles become smaller. The broadest \emph{and} most shifted profiles undergo the largest changes. The line profiles themselves are also strongly affected, as Figure~\ref{fig:NiPlumeProfiles_full} shows -- but there is no uniform pattern between the different elements. In particular for Mg~I], there is the notable outcome that the central Mg~I] emission becomes much more prominent for the aligned viewers, leading to brighter Mg~I] emission for these viewers despite the lower total energy deposition. Meanwhile, for the anti-aligned viewers we note no difference in their line profiles. The brightening of the Mg~I] profiles in the masked setup can be attributed to the loss of blocking lines in the masked region (see Figure~\ref{fig:MgFe_taucomp}), indicating that the emergent Mg~I] profile strongly depends on radiative transfer effects, with the profile as a whole thereby not necessarily appearing similar as for other lines, such as [O~I] or H$\alpha$. On the other hand, H$\alpha$ and [Ca~II] are affected in a more predictable manner, and they lose the emission specifically coming from the $^{56}$Ni plume in the masked setup but otherwise retain their profiles.} 

\txtred{\citet{jerkstrand2015late} previously studied the 1D time-dependent transmission curves of Mg~I] $\lambda\,4571$ and [O~I] $\lambda\lambda\,6300,\,6364$ and found that line blocking decreases over time but can remain significant for Mg~I] quite late into the nebular phase. It typically leads to blueshifted centroids at early times, which adjust over time towards the rest wavelength (which was also observed in stripped envelope SNe by e.g. \citealt{taubenberger2009nebular} for [O~I]). The complex behaviour of $\tau_\text{line}$ for Mg~I] and Fe~II $\lambda\,4585$ displayed in Figure~\ref{fig:MgFe_taucomp} in reaction to the masking of the $^{56}$Ni plume indicates that the temporal evolution of the line profiles alongside their line widths and centroid shifts can be used as a probe of asymmetric SN ejecta geometries, since the 3D transmission curves react differently due to the appearance of holes which can make line blockers vanish along specific sight lines. However, as the line profiles for Mg~I] in Figure~\ref{fig:NiPlumeProfiles_full} show, the Mg~I] profile itself cannot be considered a complete tracer for the geometry of the system, as the viewers anti-aligned with the primary $^{56}$Ni plume see no difference between the regular and masked model.} 

\txtred{Some parts of what we found here could be unique to this specific model, which possesses a single, clearly dominant Ni plume and which in itself is a relatively special case as a low-mass progenitor exploded with a low explosion energy. Future work will have to be done to investigate if the alignment of the Ni plume and the line centroid blueshift (correlating with angle $\xi$) also apply to other progenitors with other masses and explosion energies. Other studies have already found that prominent $^{56}$Ni plumes can be formed at higher masses \citep[e.g.][]{vartanyan2025simulated,giudici2025hydrodynamic}. If such alignments occur for (many) SNe, it would indicate tremendous potential to probe explosion asymmetries in observed SNe by gathering large samples of SNe with similar progenitors, and probing the frequency of large centroid shifts and line widths within this sample, to establish if Ni plumes are a dominant cause for line profile asymmetries. \txtred{From Figure~\ref{fig:NiPlume_shiftwidth}, it can be seen that part of the $v_\text{shift}$:$v_\text{width}$ regime can \emph{only} be achieved in a setup with a strong $^{56}$Ni plume.}}

There are two important radiative transfer mechanisms which are not yet included in \texttt{ExTraSS}: electron scattering and dust. Electron scattering will play more important roles at relatively early times when the electron scattering optical depth is higher, and can lead to a small blueshift of the peak and an enhanced red tail \citep[see e.g.][Figure~4]{jerkstrand2017spectra} \txtred{-- this could cause the extended red wing in the observed Mg~I] profile in SN~2016bkv which is not matched by our model here}. Dust on the other hand likely becomes a more important effect for later times (and was important in e.g. SN~2008bk), and will absorb (part of) the red side of line profiles and shift the peak of the profiles more strongly to the blue than for electron scattering (if $\tau_\text{dust} = \tau_\text{ES}$, and the dust is partially absorptive).

\section{Conclusions} \label{sec:conclusion}
In this work, we applied the updated 3D NLTE radiative transfer spectral synthesis code \texttt{ExTraSS} \txtred{(which is described in detail in \citealt{vanbaal2025ExTraSS})} to make spectral predictions of a $M_\mathrm{ZAMS}=9.0\,M_\odot$ star exploded in 3D by \citet{melson2020resolution} and with long-term hydrodynamical evolution by \citet{stockinger2020three}. Compared to 1D (\J18), compositional structure and $\gamma$-deposition is more spread out in 3D. The luminosities of important nucleosynthesis lines -- Mg~I] $\lambda\,4571$, [O~I] $\lambda\lambda\,6300,\,6364$, [Ca~II] $\lambda\lambda\,7291,\,7323$ and [C~I] $\lambda\,8727$ show modest changes, factor $\lesssim$2. However, luminosities of Fe~I lines and H$\alpha$ change by factors \txtred{2-3}. 

With the new radiative transfer, we show that residual opacity can affect line luminosities differently for different lines of sight. At $400\,$d the variation is $\sim$\txtred{15}\,\% for H$\alpha$, but $\sim$\txtred{60}\,\% for Mg~I] $\lambda\,4571$. This gives important information for the level of detail to which it is meaningful to analyse spectra using 1D models, and what degree of uncertainty should be accounted for when interpreting observations. The viewing angle variation also changes the centroid of lines (range \kms{\pm200}) and their FWHM (variation $\sim$50\%). We show a clear correlation for some of the lines between centroid shift and line width\txtred{, and strong correlation with respect to the Ni plume present in the model and the measured line properties}.

% Data comparison
The model spectra were compared against SN~1997D and SN~2016bkv, two low-luminosity Type IIP SNe. For SN~1997D, the model H$\alpha$ is too strong, but Mg~I] $\lambda\,4571$, [O~I] $\lambda\lambda\,6300,\,6364$, [Fe~II] $\lambda\,7155$ and [Ca~II] $\lambda\lambda\,7291,\,7323$ are reasonably well reproduced, both regarding luminosities and line profiles. This reinforces the conclusion of \J18, that $\sim0.1\,$B explosions of low-mass progenitors ($M_\mathrm{ZAMS}=8-10$ $M_\odot$) are the counterparts of LLIIP SNe -- now with validation from 3D modelling. The good match of line strengths and profiles, for many lines, is encouraging for the \txtred{neutrino}-driven Fe-CCSN paradigm as explosion mechanism. However, more such 3D models with more complete physics are needed -- for example, a larger nuclear network for the nucleosynthesis calculations. Our model here has lower explosion energy than inferred from a sample of LLIIP SN light curves \citep{das2025low}, although explosions with such low energies can still lead to good matches of multi-band light curves for some events \citep{kozyreva2022low}, and we find good matches to the line profiles of SN~1997D and SN~2016bkv here.

\txtred{Through masking the $\gamma$-rays originating from the primary $^{56}$Ni plume present in the model, we also study the impact this plume has on the line profiles and find that the properties of the line profiles strongly correlate with this primary $^{56}$Ni plume in the regular setup. Furthermore, we find that the masked line profiles react differently for various elements, with Mg~I] $\lambda\,4571$ actually brightening for some viewers, while H$\alpha$ and [Ca~II] $\lambda\lambda\,7291,\,7323$ become dimmer, as various line blockers for Mg~I] disappear from the plume region, enabling more central emission to leak out to these observers.}

An important goal of investigating LLIIP SNe with spectral models is to search for electron-capture supernova signatures. A key result here is that the prediction using 1D models of which lines distinguish an Fe-CCSN from an ECSN -- unique He~I $\lambda\,7065$, [C~I] $\lambda\,8727$, [C~I] $\lambda\lambda\,9824,\,9850$, and stronger Mg~I] $\lambda\,4571$, [O~I] $\lambda\lambda\,6300,\,6364$ (\J18) -- are upheld in 3D (an additional distinguishing line is O~I 8446, however the effects responsible for creating this line are not simulated here). While in 3D there is no "wall" of nucleosynthesis layers as in 1D, and thus the $\gamma$-ray energy deposition profile is different, the change in line luminosities is moderate enough (factor $\lesssim$ 2) that these lines are still distinct. SN~2016bkv therefore remains as one of the strongest candidates for an ECSN \citep{hosseinzadeh2018short} -- showing no clear lines from hydrostatic He layers. The main discrepancy is however an estimated $^{56}$Ni mass that is a factor several above what ECSNe are expected to produce.

With the inclusion of UVOIR radiative transfer into \texttt{ExTraSS}, it has become possible to investigate the finer details of line profile properties. Photoionization and photoexcitation are particularly important for Type II SNe in the nebular phase \citep{jerkstrand2012progenitor}. With the large grid of 3D hydrodynamic models available for such explosions \citep[e.g.][and more models in the pipeline]{stockinger2020three,gabler2021infancy,vartanyan2025simulated}, it will be of interest to study the spectral signatures of these models to compare to the whole population of Type II SNe. Detailed line profile analysis with \texttt{ExTraSS} can be used to constrain the explosion energies of the SNe, and thus reduce uncertainties that are connected to the light curve analysis.
% Don't you want to mention the capability to determine/constrain explosion energies of SNe with the line analysis enabled by ExTraSS, and thus to reduce uncertainties connected to the light-curve analysis? 

% The last numbered section should briefly summarize what has been done, and describe the final conclusions which the authors draw from their work.

\section*{Acknowledgements}

\txtred{The authors would like to thank the referee for their comments and suggestions, and} to thank Tobias Melson and Georg Stockinger for their work on the original explosion modelling used for this work. 
BvB would like to express his deepest gratitude to Arash Alizad Banaei for helping with resolving the computational issues between Dardel and \texttt{ExTraSS}.
The authors acknowledge support from the European Research Council (ERC) under the European Union’s Horizon 2020 Research and Innovation Programme (ERC Starting Grant No. [803189], PI: A. Jerkstrand). The computations were enabled by resources provided by the National Academic Infrastructure for Supercomputing in Sweden (NAISS) at the PDC Center for High Performance Computing, KTH Royal Institute of Technology, partially funded by the Swedish Research Council through grant agreement no. 2018-05973. This work was supported by the German Research Foundation (DFG) through the Collaborative Research Centre ``Neutrinos and Dark Matter in Astro- and Particle Physics (NDM),'' Grant SFB-1258$\,$--$\,$283604770, and under Germany's Excellence Strategy through the Cluster of Excellence ORIGINS EXC-2094$\,$--$\,$390783311.

%%%%%%%%%%%%%%%%%%%%%%%%%%%%%%%%%%%%%%%%%%%%%%%%%%
\section*{Data Availability}

The Data underlying this article will be shared on reasonable request to the corresponding author.

\noindent The nebular phase spectra included in this work were all obtained from \href{https://www.wiserep.org}{WiSeREP} -- \citet{wiserep2012}.



\bibliographystyle{mnras}
\bibliography{bibliography}

%\begin{thebibliography}{99}
%\end{thebibliography}

%%%%%%%%%%%%%%%%%%%%%%%%%%%%%%%%%%%%%%%%%%%%%%%%%%

%%%%%%%%%%%%%%%%% APPENDICES %%%%%%%%%%%%%%%%%%%%%

\appendix

\section{\txtred{250 day line profiles}}  \label{app:OI250day}
\begin{figure*}
    \centering
    \includegraphics[width=\linewidth]{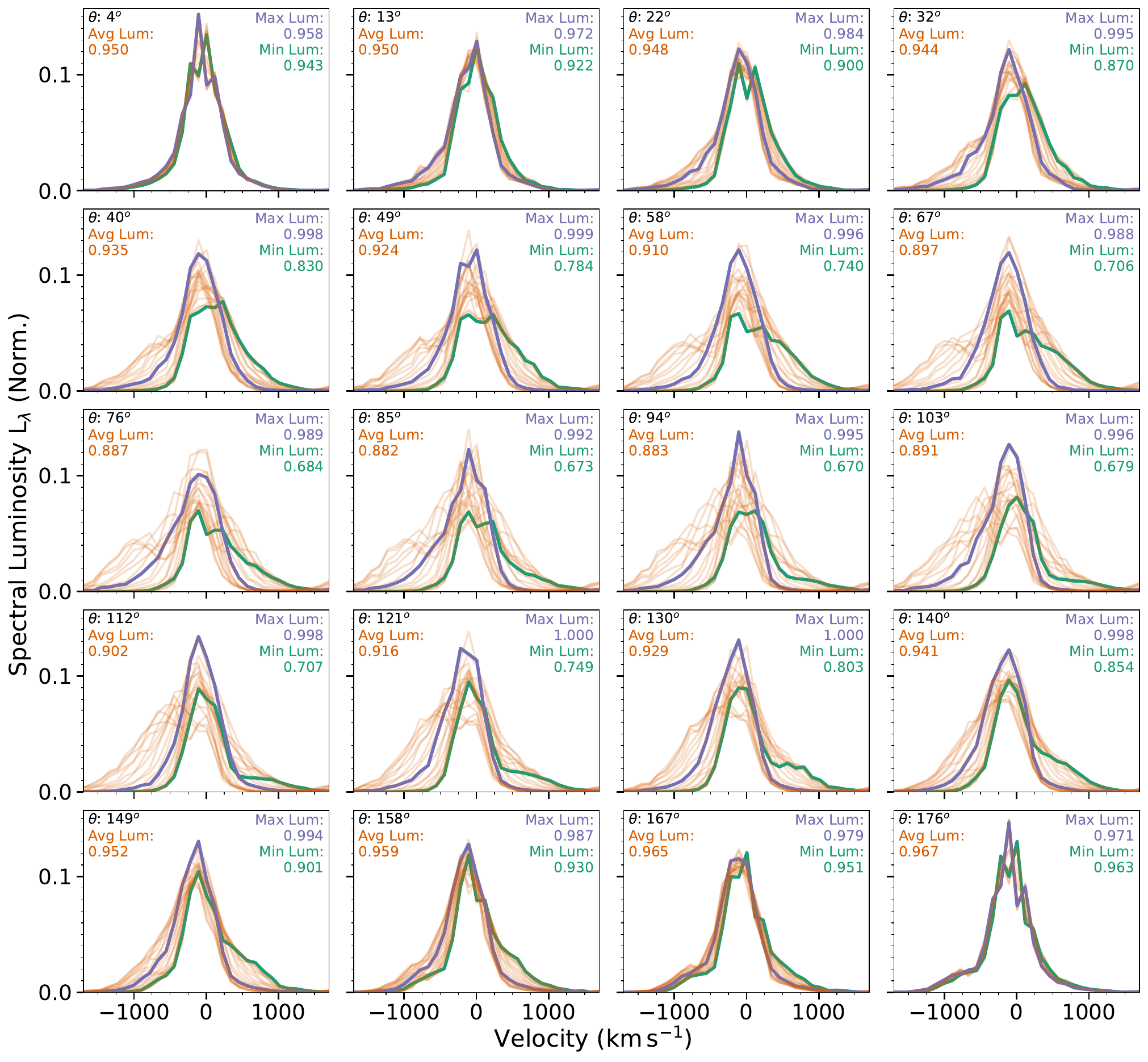}
    \caption{\txtred{The [O~I] line profiles in s9.0 at 250 days, similar as in Figure~\ref{fig:OI-400_fullVA}, with the normalization of the profiles accounting for both [O~I] features. For clarity, we only show the $6300\,\angstrom$ feature in this plot. The most luminous profile is the purple profile at $\theta=121^\circ$.}}
    \label{fig:OI-250_fullVA}
\end{figure*}
\txtred{In this appendix, we show the $20\times20$ line profiles for [O~I] $\lambda\,6300$ at 250 days post-explosion, similar as in Figure~\ref{fig:OI-400_fullVA}. For clarity, the [O~I] $\lambda\,6364$ component is not shown, as the two lines are fully separated.}

% \section{\txtred{Line profile properties towards $^{56}$Ni plume}} \label{app:NiPlumeApp}
% \begin{figure*}
%     \centering
%     \includegraphics[width=.495\linewidth]{Figures/Stock9-400_combo_Mg-4571_v1800-NiPlume.pdf}
%     \includegraphics[width=.495\linewidth]{Figures/Stock9-400_combo_O-6300_v1500-NiPlume.pdf}
%     \includegraphics[width=.495\linewidth]{Figures/Stock9-400_combo_H-6565_v1800-NiPlume.pdf}
%     \includegraphics[width=.495\linewidth]{Figures/Stock9-400_combo_Fe-7157_v1800-NiPlume.pdf}
%     \includegraphics[width=.495\linewidth]{Figures/Stock9-400_combo_Ca-7304_v2585-NiPlume.pdf}
%     \includegraphics[width=.495\linewidth]{Figures/Stock9-400_combo_C-8727_v1800-NiPlume.pdf}
%     \caption{\txtred{The same as Figure~\ref{fig:profile_props}, but colour coded by the angle $\xi$, which is the angle between the observer and the direction of the main $^{56}$Ni plume (red cross in Figure~\ref{fig:Niplume-aitoff}). $\xi = 0$ means that the centre of the Ni plume points exactly to the observer.}} 
%     \label{fig:NiPlume_props}
% \end{figure*}
% \txtred{In this appendix, we show the line profile properties for the main features (Mg~I], [O~I], H$\alpha$, [Fe~II], [Ca~II] and [C~I]) are shown, colour coded by $\xi$. The calculations of $v_\text{shift}$ and $v_\text{width}$ have not been changed, so in Figure~\ref{fig:NiPlume_props} only the colours vary compared to Figure~\ref{fig:profile_props}.}

%%%%%%%%%%%%%%%%%%%%%%%%%%%%%%%%%%%%%%%%%%%%%%%%%%

% Don't change these lines
\bsp	% typesetting comment
\label{lastpage}
\end{document}